\documentclass[twocolumn]{aastex7}
\usepackage{CJKutf8}
\graphicspath{{./},{./figures/}}
\usepackage{amsmath}
\usepackage{mathptmx,txfonts}
\usepackage{booktabs}
\usepackage{xspace}
\usepackage{threeparttable}
\usepackage{bm}
\usepackage{tablefootnote}
\usepackage[capitalize]{cleveref}

\newcommand{\numax}{{\ensuremath{\nu_{\text{max}}}}}
\newcommand{\pbjam}{{\texttt{PBJam}}\xspace}
\newcommand{\reggae}{{\texttt{reggae}}\xspace}

\begin{document}

\correspondingauthor{Christopher Lindsay}
\email{christopher.lindsay@yale.edu}

\title{The Effect of Different Methods for Accounting for $\alpha$-enhancement on the Asteroseismic Modeling of Metal-Poor Stars}

\author[0000-0001-8722-1436,gname='Christopher',sname='Lindsay']{Christopher J. Lindsay}
\affiliation{Department of Astronomy, Yale University, PO Box 208101, New Haven, CT 06520-8101, USA}
\email{christopher.lindsay@yale.edu}

\author[0000-0001-7664-648X,gname='Joel',sname='Ong']{J. M. Joel Ong\begin{CJK*}{UTF8}{gbsn}(王加冕)\end{CJK*}}
\affiliation{Institute for Astronomy, University of Hawai`i, 2680 Woodlawn Drive, Honolulu, HI 96822, USA}
\affiliation{Hubble Fellow}
\email{joelong@hawaii.edu}

\author[0000-0002-6163-3472,gname='Sarbani',sname='Basu']{Sarbani Basu}
\affiliation{Department of Astronomy, Yale University, PO Box 208101, New Haven, CT 06520-8101, USA}
\email{sarbani.basu@yale.edu}

\author[0000-0003-4976-9980, gname='Samuel',sname='Grunblatt']{Samuel Grunblatt}
\affiliation{Department of Physics and Astronomy, University of Alabama, 514 University Drive, Tuscaloosa, AL 35487-0324, USA}
\email{sgrunblatt@ua.edu}

\author[0000-0003-2400-6960, gname='Marc',sname='Hon']{Marc Hon}
\affiliation{Department of Physics, National University of Singapore, 21 Lower Kent Ridge Road, Singapore, 119077}
\affiliation{Kavli Institute for Astrophysics and Space Research, Massachusetts Institute of Technology, Cambridge, MA 02139, USA}
\email{mtyhon@nus.edu.sg}

\shortauthors{Lindsay, et al.}
\shorttitle{$\alpha$-enhanced Stellar Modeling}

\begin{abstract}
Constraining stellar models using asteroseismic and spectroscopic observations is a powerful method for precisely determining the fundamental properties of stars in different kinematic components of our galaxy. We use spectroscopy and individual oscillation mode frequencies to perform a homogeneous modeling study of eight evolved metal-poor stars enhanced in $\alpha$-elements. We compare a full treatment of $\alpha$-enhancement against an ad hoc correction to the total metallicity and show that the stellar properties inferred from asteroseismic modeling using both sets of models agree with each other. Additionally, we find that the uncertainties on stellar parameters derived from the both $\alpha$-enhanced modeling methods are comparable. This is in qualitative disagreement with existing works showing red-giant ages constrained by only the global asteroseismic parameters to depend strongly on the opacities and abundances assumed in 1D modeling. We also show that the observed frequency of maximum oscillation power ($\nu_{\text{max}}$) is larger than the value predicted from applying the $\nu_{\text{max}}$ scaling relation to the masses, radii, and temperatures inferred from the detailed modeling. This discrepancy is pronounced at low metallicities, consistent with recent findings indicating a breakdown of the $\nu_{\text{max}}$ scaling relation for metal-poor stars. Understanding the extent to which the $\nu_{\text{max}}$ scaling relation fails for low-metallicity solar-like oscillators through detailed modeling will enable more accurate mass and age determinations for hundreds of giant stars in the Galactic Halo for which only global asteroseismic parameters are available.
\end{abstract}

\keywords{asteroseismology - stars: solar-type - stars: oscillations - Galaxy: halo - Galaxy: formation}

\section{Introduction} \label{sec:intro}
Combined position and velocity measurements from \textit{Gaia} \citep{Gaia_mission, Gaia_DR2, Gaia_DR3}, and spectroscopic abundance data from experiments like APOGEE \citep{apogee}, have enabled the detailed study of substructure in the Milky Way's Halo. This rich substructure in turn evinces the complex, hierarchical history of structure formation in the Milky Way. In order to determine when merger events occurred, and study the properties of the smaller galaxies that merged with the Milky Way, it is vital to determine ages for stars in the Halo. The age-dating of halo stellar populations associated with merger events has been carried out extensively for this purpose by comparing the spectroscopic qualities of stars in these populations to grids of stellar isochrones \citep[e.g.][]{Koppelman2019, XiangRix2022, Ruiz-Lara_helmi_history, Gallart2024}. 

Asteroseismic observations taken by space-based missions --- such as CoRoT \citep{Baglin2006}, \textit{Kepler} \citep{Kepler_inst} and TESS \citep{TESS_inst} --- have previously been used, together with stellar spectra, to place precise constraints on the fundamental stellar parameters of thousands of stars \citep[e.g.][]{APOKASC2, APOKASC3, apok2, Marasco2025}. Stars with outer convective envelopes, like our Sun, subgiants, and evolved giant stars, pulsate in many oscillation modes simultaneously. The stochastically-excited oscillations of these `solar-like' oscillators produce peaks in power spectra that are obtained by taking the Fourier transform of a star's light curve \citep[see][for a recent review of giant star asteroseismology]{2017A&ARv..25....1H}. These oscillation peaks produce a comb-like pattern of regularly spaced pressure modes since modes with the same angular degree, $\ell$, show a consistent frequency spacing between overtones called the large frequency separation, $\Delta \nu$ \citep{Tassoul1980}. $\Delta \nu$ scales with the square root of the stellar density giving the $\Delta \nu$ scaling relation
\begin{equation}
    \Delta \nu \approx \Delta \nu_{\odot} \sqrt{\frac{M/M_{\odot}}{(R/R_{\odot})^{3}}}. 
\end{equation} 

The oscillation peaks do not all have the same strength; instead, they appear under a Gaussian-like envelope with the peak of the envelope, called the frequency of maximum oscillation power. This frequency, $\nu_{\text{max}}$, scales with the star’s surface gravity and temperature \citep{Brown1991, Kjeldsen1995} resulting in the $\nu_{\text{max}}$ scaling relation, given by
\begin{equation}
    \label{eq:numax_scaling}
    \nu_{\text{max}} \approx \nu_{\text{max}, \odot} \frac{M/M_{\odot}}{(R/R_{\odot})^{2}}\sqrt{\frac{T_{\text{eff},\odot}}{T_{\text{eff}}}}
\end{equation} 

Using these global asteroseismic scaling relations, measurements of temperature, $\Delta \nu$, and $\nu_{\text{max}}$ directly return stellar masses and radii. Combining asteroseismic mass and radius determinations with stellar models permits more precise field star age determinations than are possible without the use of asteroseismology \citep{soderblom_1, soderblom_2}. When individual mode frequencies can be determined from observed stellar power spectra, comparing the observed mode frequencies with theoretical oscillation mode frequencies predicted by stellar models provides constraints on stellar interior structure and enables more precise determinations of global parameters such as age.

Asteroseismic Galactic archaeology studies on a large scale have previously been carried out using global asteroseismic measurements derived from \textit{Kepler} or CoRoT data \citep[e.g.][]{Silva_Aguirre2018, Miglio2021, warfield, apok2, Stokholm2023}, and now are done using the large influx of data from TESS \citep[e.g.][]{Grunblatt2021, Borre2022, Marasco2025}. The asteroseismic analyses of stars in different kinematic components of the Milky Way have thus far primarily relied on the global asteroseismic scaling relations to determine stellar masses and radii, and thereby ages through stellar models. However, recent work indicates that scaling relation-derived stellar parameters diverge from detailed asteroseismic modeling results, obtained by fitting the observed mode frequencies to the mode frequencies of stellar models \citep{Huber2024, Larsen2025, Lindsay2025}. 

Accurate and precise stellar ages for metal-poor stars in the Galactic Halo are vital for determining the merger history of our galaxy. Using a $\Delta \nu$ measurement determined from radial velocity measurements of the bright metal-poor Halo star $\nu$ Indi \citep{Carrier2007}, \citet{Bedding2006} placed a lower bound of 9 Gyr on the age of the star. Later, a precise age determination for $\nu$ Indi \citep[by][]{Chaplin2020} using individual mode frequencies established with TESS photometric data placed a limit on the timing of the Gaia-Enceladus merger event, which is largely responsible for the formation of the inner Galactic Halo and Thick Disk \citep{Helmi2018}. More recently, \cite{Lindsay2025} also used individual mode frequency asteroseismic modeling to place a lower bound on when stars began to form in the dwarf galaxy progenitor of the Helmi streams \citep{Helmi1999}. Asteroseismology with individual mode frequencies remains the gold standard for determining precise stellar ages. However, modeling uncertainties such as the structural resolution of the models, mixing near convective boundaries, and treatments for non-solar elemental mixtures must now be explored \citep{LiJoyce2025, Lindsay2024}. 

\begin{figure*}[ht!]
    \centering
    \includegraphics[width=0.45\textwidth]{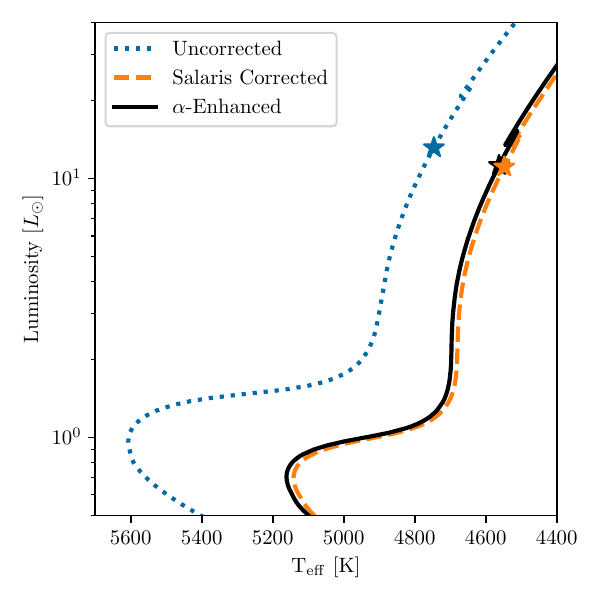}
    \includegraphics[width=0.45\textwidth]{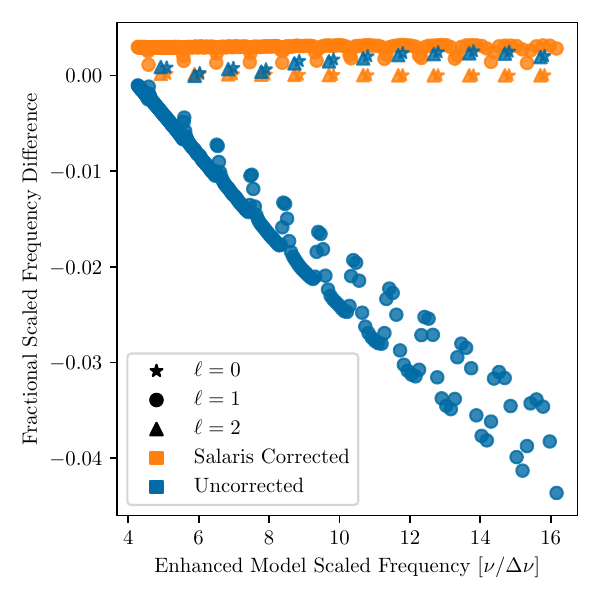}
    \includegraphics[width=0.45\textwidth]{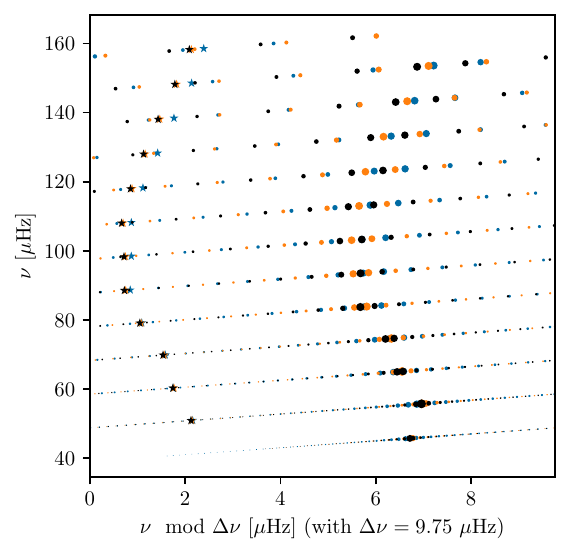}
    \includegraphics[width=0.45\textwidth]{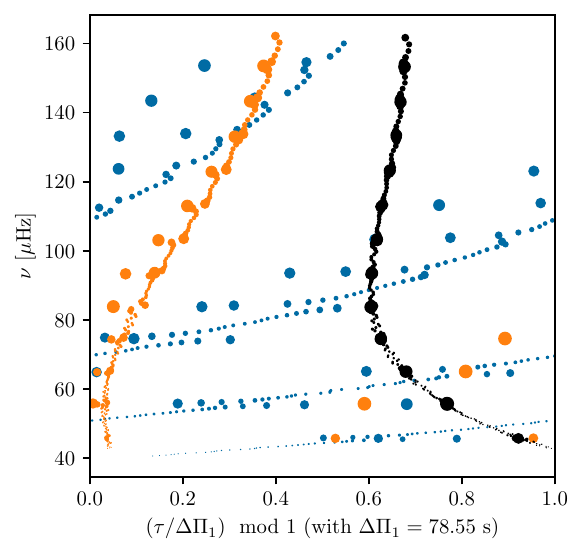}
    \caption{Top Left: HR-Diagram evolutionary tracks, showing how stellar evolution changes when modeling stars that are enhanced in $\alpha$-elements. The uncorrected track (blue dotted line) shows the evolution of a 0.85 $M_{\odot}$ star with initial helium abundance $Y_0 = 0.276$ and initial iron abundance [Fe/H]$_0$ = 0.0. The $\alpha$-enhanced track (black line) has the same values of mass, $Y_0$, and [Fe/H]$_0$ as the uncorrected track but incorporates an $\alpha$-enhancement factor of [$\alpha$/Fe] = 0.4. The Salaris-corrected track (orange dashed line) has a different [Fe/H]$_0$ value than the uncorrected track and instead matches the initial global metallicity ($Z_0$) of the $\alpha$-enhanced track, while keeping the same solar-scaled \citep{GS98} element abundance mixtures as the uncorrected track. Top Right: Model mode frequency comparison between the three models with the same acoustic radius such that $\Delta\nu \sim 1/2T = 10\ \mu$Hz, marked with stars in the left panel. The scaled frequencies (mode frequency divided by $\Delta \nu$) are plotted on the x-axis while the scaled frequency difference (Salaris-corrected or uncorrected scaled frequency minus the $\alpha$-enhanced model scaled frequency) divided by the $\alpha$-enhanced model scaled frequencies are plotted on the y-axis. The radial (star symbols) and quadrupole (triangle symbol) mode scaled frequency differences are small for both the Salaris-corrected and uncorrected models, however the dipolar mode scaled frequency difference (circle symbols) are large when comparing the uncorrected and $\alpha$-enhanced models. Bottom Left: Comparison of frequency-echelle diagrams for the three models shown. Symbols denoting angular degree are colored in the same way as the top right panel, and are sized by the mixing fractions $\zeta$ --- larger symbols denote more p-like modes and smaller ones denote g-like modes. The p-mode frequencies for the models are similar, as shown by the locations of the radial modes and of the most p-dominated dipole mixed modes. However, the exact configuration of the dipole mixed-mode pattern is very different. Bottom Right: Stretched period-echelle diagrams for the dipole modes of the three models, constructed using the period spacing, coupling strengths, and pure p-mode frequencies of the $\alpha$-enhanced model. The three models can clearly be seen to have very different period spacings $\Delta\Pi_{\ell=1}$, as well as different g-mode phase offsets $\epsilon_g$, different pure p-mode frequencies, and different coupling strengths $q$ between the p- and g-modes.}
    \label{fig:alpha_methods}
\end{figure*}

In this work, we explore the effects of using different ways of treating $\alpha$-element enhancement on asteroseismic stellar modeling. Metal-poor Halo stars are often enhanced in $\alpha$ elements compared with the Sun. These $\alpha$ elements form predominantly through the $\alpha$-process and have mass numbers that are multiples of 4, including C, O, Ne, Mg, etc. They are so enhanced due to the early metal enrichment of the interstellar medium being the result of core-collapse supernovae, which produce higher concentrations of $\alpha$-elements \citep{Spite1978, Tinsley1980, McWilliam1995, Nissen2010, Kobayashi2020}. Therefore, the Milky Way's old stars are enhanced in these $\alpha$-elements when compared with younger stars like our Sun. Despite this, $\alpha$-enhanced stars are often not generally studied using models that account for deviations from Solar elemental abundance patterns. Instead, the stars are often modeled using solar-scaled metallicities, with an empirical correction applied to the measured global metallicity \citep{salaris},
\begin{equation}
    \label{eq:salaris_correction}
    \textrm{[Fe/H]}_{\textrm{corr}} \simeq \textrm{[Fe/H]}_{\textrm{orig}} + \log_{10}(0.638 \times 10^{\textrm{[$\alpha$/Fe]}} + 0.362).
\end{equation}
The use of \autoref{eq:salaris_correction} (which we will subsequently call the Salaris correction) allows for the construction of stellar evolutionary tracks that match a star's position in the HR diagram. \autoref{fig:alpha_methods} illustrates this by comparing how evolutionary tracks of stellar models, calculated with the MESA stellar evolution code \citep{Paxton2011, Paxton2013, Paxton2015, Paxton2018, Paxton2019, Jermyn2023}, respond to these treatments of $\alpha$-enhancement \citep[see also][]{Sun2023}. At 0.85 solar masses, the $\alpha$-enhanced track (black solid line in the left panel of \autoref{fig:alpha_methods}), calculated using $\alpha$-enhanced element mixtures and opacity tables, lies at a lower temperature compared with the uncorrected track (blue dotted line in the left panel of \autoref{fig:alpha_methods}). Although the $\alpha$-enhanced and uncorrected track both have the same initial iron abundance of [Fe/H]$_0$ = 0.0, the initial global metal mass abundances ($Z_0$) differ due to the different elemental mixture. The Salaris-corrected track (orange dashed line in the left panel of \autoref{fig:alpha_methods}), calculated using a solar-scaled element mixture and solar-scaled opacity tables, mostly coincides with the $\alpha$-enhanced track since it has a different [Fe/H]$_0$, but exactly the same $Z_0$.

The internal structure of such an $\alpha$-enhanced stellar model may not necessarily match that of an equivalent Salaris-corrected model constructed using solar-scaled abundances and opacities. As a result, the detailed asteroseismic properties and mode frequencies of $\alpha$-enhanced and Salaris-corrected models may not reproduce each other, either. 

In this work, we seek to examine potential $\alpha$-enrichment asteroseismic modeling systematic effects more closely. We build on previous studies of $\alpha$-enhanced asteroseismic stellar modeling by conducting a homogeneous modeling study of 8 metal-poor, $\alpha$-enhanced stars, separately fitting $\alpha$-enhanced and Salaris-corrected stellar models to each star based on combined constraints from spectroscopic observables and individual oscillation mode frequencies. Our goal in this work is to determine whether applying a full treatment of $\alpha$-enhancement (altering abundance fractions and using opacity tables) significantly influences the stellar parameters estimated from asteroseismic modeling. We begin by illustrating the differences in mode frequencies between $\alpha$-enhanced and non-$\alpha$-enhanced stellar models in \autoref{sec:mode_differences} We introduce our sample of 8 metal-poor, $\alpha$-enhanced stars we study in the work in \autoref{sec:sample} and describe the sources of the spectroscopic and asteroseismic data used in the modeling. We detail our asteroseismic modeling methods in \autoref{sec:optimization}. The modeling results obtained using the different methods of accounting for $\alpha$-enhancement are compared in \autoref{sec:results} and implications of the modeling results on metal-poor, $\alpha$-enhanced asteroseismic stellar modeling are discussed in \autoref{sec:discussion}.

\section{Model Mode Differences}
\label{sec:mode_differences}
We illustrate differences in the mode frequencies between $\alpha$-enhanced and non-$\alpha$-enhanced stellar models with the same global asteroseismic parameters in the right panel of \autoref{fig:alpha_methods}. We calculate the radial and quadrupole pressure modes, as well as the dipolar mixed modes for three three models with the same acoustic radius ($T$). The three models (uncorrected, Salaris-corrected, and $\alpha$-enhanced) are taken so that $1/(2T) = 10\,\mu\mathrm{Hz} \sim \Delta\nu$, and the position of these models are shown with stars in the evolutionary tracks plotted in the upper left panel of \autoref{fig:alpha_methods}. Comparing their scaled frequencies (mode frequency divided by the $\Delta \nu$ of each model) we can see that the fractional scaled frequency difference between the modes calculated from the Salaris-corrected model and the $\alpha$-enhanced model (orange points in the upper right panel of \autoref{fig:alpha_methods}) are small between the two models with the same initial metal mass fraction, $Z_0$. In comparison the fractional scaled frequency difference between the modes calculated from the uncorrected model and the $\alpha$-enhanced model (blue points in the upper right panel of \autoref{fig:alpha_methods}) are larger, particularly for the dipolar ($\ell = 1$) modes. For both comparisons, the fractional scaled frequency differences are larger for the dipole modes (marked in circles) compared with the differences calculated using the radial modes (star symbols) or quadrupole modes (triangle symbols) since the dipole modes are mixed modes which are more sensitive to the properties of the stellar core and depend more strongly on the wether or not $\alpha$-enhanced opacity tables are used to calculate the stellar models.

These differences can also be visualized in frequency- or period-echelle diagrams, which we show in the lower two panels of \autoref{fig:alpha_methods}. In constructing these, we have adopted the large separation $\Delta\nu$ and dipole g-mode period spacing $\Delta\Pi_1$ of the model calculated with $\alpha$-enhanced opacity tables. The period-echelle diagram shown in the lower right panel of \autoref{fig:alpha_methods} is stretched to disentangle the pure g-mode frequencies from coupling with p-modes, and is constructed using the same set of coupling strengths $q$ and p-mode frequencies ($\nu_p$) for all three models, following the procedure outlined in \citet{ong_mode_2023}. Since these models are selected to possess similar acoustic radii the location of the more p-mode dominated dipolar mixed modes can be seen to lie in the same location on the frequency-echelle diagram. However, the period-echelle diagrams show very different structures. Calculating models using $\alpha$-enhanced opacity tables results in a slightly larger dipolar period spacing (the characteristic separation in period between consecutive dipolar gravity modes, $\Delta\Pi_{\ell = 1}$) compared to models using solar-abundance based opacity tables. The MESA-calculated $\Delta\Pi_{\ell = 1}$ for the $\alpha$-enhanced model shown in \autoref{fig:alpha_methods} is 79.2s while $\Delta\Pi_{\ell = 1} = 78.8$s for the Salaris-corrected model and $\Delta\Pi_{\ell = 1}=77.9$s for the uncorrected model.

Structural differences may in turn lead to discrepancies in stellar properties inferred from asteroseismology using Salaris-corrected models compared to those obtained using properly $\alpha$-enhanced ones. Some existing works suggest that these discrepancies may be nontrivial. For example, \citet{Ge2015} performed a grid-based modeling study of the evolved subgiant KIC7976303 and found the $\alpha$-enhanced models fit the individual observed frequencies better than models constructed without $\alpha$-enhancement, and that $\alpha$-enhanced modeling implied a larger mass and younger age for the star compared with non-$\alpha$-enhanced modeling. On the other hand, \citet{Ge2015} also found that both $\alpha$-enhanced and solar-scaled models matched the observations of KIC8694723. \citet{Valle2024} also found that using solar-scaled abundances and opacities in their grid-based stellar modeling led to biased ages. However, this latter work only used the global asteroseismic parameters. It is difficult to assess how important $\alpha$-enhanced modeling might be for asteroseismic inference by generalizing only this handful of studies, which is why we perform detailed asteroseismic modeling for the stars in our sample.

\section{The Sample}
\label{sec:sample}

We perform detailed asteroseismic modeling on a sample of 8 metal-poor stars, all of which are enhanced to varying degrees in $\alpha$-elements relative to the Sun. We report the values of effective temperature (T$_\text{eff}$), [Fe/H], [$\alpha$/Fe], and luminosity ($L$) we use in our modeling in \autoref{table:spec_inputs}, along with their associated sources. In order to ensure consistency between the stars in our sample, the [$\alpha$/Fe] measurements are determined by taking a weighted average of the elemental abundances of Mg, Si, Ca, and Ti cited in \autoref{table:spec_inputs}. Each [$X$/Fe] value is weighted by one over the squared error following \citep[e.g.][]{Lindsay2025, Lundkvist2025}.

\begin{figure*}
    \centering
    \includegraphics[width=0.46\textwidth]{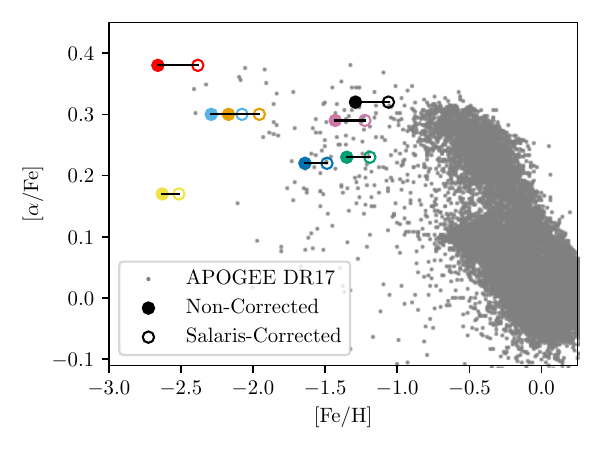}
    \includegraphics[width=0.46\textwidth]{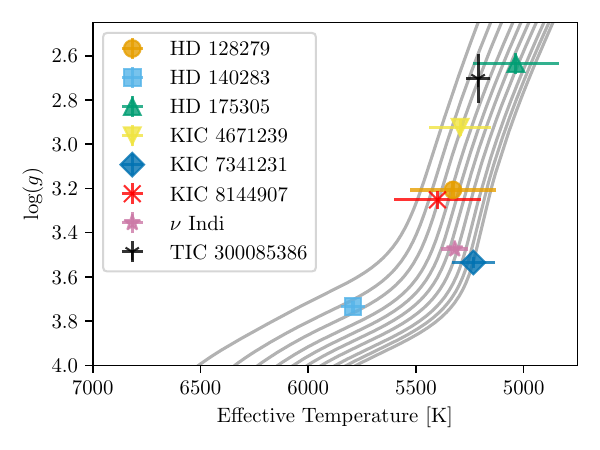}
    \caption{Left Panel: The observed [Fe/H] (filled symbols) and Salaris-corrected [Fe/H] (open symbols) values against the [$\alpha$/Fe] measurements for the stars in our sample. The background small points show the distribution of stars in the APOKASC-3 sample in the [$\alpha$/Fe]-[Fe/H] plane. The filled points show the [Fe/H] values for HD 128279 (orange), HD 140283 (light blue), HD 175305 (green), KIC 4671239 (yellow), KIC 7341231 (dark blue), KIC 8144907 (red), $\nu$ Indi (pink), and TIC 300085386 (black), while the open symbols show the [Fe/H] values after the correction from \citet{salaris} is applied. These values are also listed in \autoref{table:spec_inputs}. Right Panel: The colored symbols show the seismic log($g$) value and effective temperature values of the stars in our sample. The error bars show the observed temperature errors as well as the seismic log($g$) errors derived from the errors in $\nu_{\text{max}}$ and T$_{\text{eff}}$. Note the seismic log($g$) errors are multiplied by 5 for visibility. The background evolutionary tracks show 0.8$M_{\odot}$ tracks with varying initial metal mass abundance values from $Z_0 = 0.0005$ to $Z_0 = 0.005$ in steps of 0.0005.  } 
    \label{fig:sample_observed_feh}
\end{figure*}

The global asteroseismic parameters, $\Delta \nu$ and $\nu_{\text{max}}$, are also reported in \autoref{table:spec_inputs}, although they are not directly included in our modeling procedure (see \autoref{sec:optimization}). The observed [Fe/H] and [$\alpha$/Fe] measurements show that the stars in our sample are all metal poor ([Fe/H] $\lesssim -1.5$) and enhanced in $\alpha$ elements (0.15 $\lesssim$ [$\alpha$/Fe] $\lesssim$ 0.4). The left panel of \autoref{fig:sample_observed_feh} shows our sample of stars as colored points in [Fe/H] versus [$\alpha$/Fe] space, with the background points showing the observed [$\alpha$/Fe] and [Fe/H] APOGEE DR17 \citep{APOGEE_DR17} measurements taken from the APOKASC-3 sample \citep{APOKASC3}. The solid-colored points show the observed [Fe/H] measurements listed in \autoref{table:spec_inputs}, while the empty points show the [Fe/H] values after applying \autoref{eq:salaris_correction}. 

The right panel of \autoref{fig:sample_observed_feh} shows the seismic $\log(g)$ and effective temperature values for the stars in our sample plotted in a Kiel diagram along with a set of 0.8$M_{\odot}$ evolutionary tracks with different metal mass abundance values varying from $Z_0 = 0.0005$ (highest temperature track) to $Z_0 = 0.005$ (lowest temperature track) in steps of 0.0005. The seismic $\log(g)$ values are calculated from the observed $\nu_{\text{max}}$ and T$_{\text{eff}}$ values using the following scaling relation, 
\begin{equation}
    \log(g)_{\text{seis}} = \log(g)_{\odot} + \log_{10}\big(\frac{\nu_{\text{max}}}{\nu_{\text{max},\odot}}\big) + 0.5 \log_{10}\big(\frac{\text{T}_{\text{eff}}}{\text{T}_{\text{eff},\odot}}\big),
\end{equation}
Adopting $\log(g)_{\odot} = 4.438$, $\nu_{\text{max},\odot} = 3090\mu$Hz and T$_\text{eff} = 5778$K, we estimate the error on $\log(g)_{\text{seis}}$ by perturbing the observed $\nu_{\text{max}}$ and T$_{\text{eff}}$ values 10000 times assuming a normal distribution with the center at the parameter value and $\sigma$ equal to the listed errors in \autoref{table:spec_inputs}. After plugging these perturbed $\nu_{\text{max}}$ and T$_{\text{eff}}$ into the equation for $\log(g)_{\text{seis}}$, we take the mean and standard deviation of the resultant distribution to be the $\log(g)_{\text{seis}}$ value and uncertainty. We obtain $\log(g)_{\text{seis}}$ = 3.207 $\pm$ 0.008 (HD 128279), 3.735 $\pm$ 0.006 (HD 140283), 2.635 $\pm$ 0.009 (HD 175305), 2.924 $\pm$ 0.008 (KIC 4671239), 3.535 $\pm$ 0.005 (KIC 7341231), 3.252 $\pm$ 0.008 (KIC 8144907), 3.474 $\pm$ 0.003 ($\nu$ Indi), and 2.704 $\pm$ 0.022 (TIC 300085386) and include these results in the right panel of \autoref{fig:sample_observed_feh}.

\begin{table*}[ht!]
\begin{threeparttable}
\caption{Spectroscopic and global asteroseismic observed values (with sources) of the 8 metal-poor stars in our sample. In cases where [$\alpha$/Fe] were not reported explicitly in the literature, we determine [$\alpha$/Fe] by taking an average of the observed element abundances of Mg, Si, Ca, and Ti, weighted by the reported uncertainties. When luminosities from Gaia DR2 \citep{Gaia_DR2} are used, we adopt a conservative 15\% error on $L$ due to the negligibly small reported errors from the pipeline and strong degeneracy between effective temperature and reddening when using the Gaia broad band photometry.  }
\label{table:spec_inputs}
\begingroup
\small
\setlength{\tabcolsep}{2pt} 
\renewcommand{\arraystretch}{1.05}
\par\noindent{\begin{tabular}{lccccccccc}
\toprule
Target & T$_{\text{eff}}$ [K] & [Fe/H] & [$\alpha$/Fe] & [Fe/H]$_{\text{corr}}$ & $L$ [$L_{\odot}]$& $\Delta \nu$ [$\mu$Hz]  & $\nu_{\text{max}}$ [$\mu$Hz] & $\Delta \Pi_{\ell = 1}$ \\
\hline
HD 128279 & $5328 \pm 200$\textsuperscript{a} & $-2.17 \pm 0.12$\textsuperscript{a} & 0.30\textsuperscript{a} & $-1.96 \pm 0.12$ & $11.03 \pm 1.124$\textsuperscript{b} & $15.87 \pm 0.05$\textsuperscript{b} & $189.10 \pm 0.5$\textsuperscript{b} & $83.5\pm0.5$\textsuperscript{b}\\
HD 140283 & $5792 \pm 55$\textsuperscript{c} & $-2.29 \pm 0.14$\textsuperscript{c} & 0.30\textsuperscript{d} & $-2.08 \pm 0.14$ & $4.77 \pm 0.055$\textsuperscript{c} & $39.47 \pm 0.05$\textsuperscript{d} & $611.30 \pm 7.4$\textsuperscript{d} & 147\textsuperscript{d}\\
HD 175305 & $5036 \pm 200$\textsuperscript{a} & $-1.35 \pm 0.15$\textsuperscript{a} & 0.23\textsuperscript{a} & $-1.19 \pm 0.15$ & $33.10 \pm 3.0$\textsuperscript{b} & $5.89 \pm 0.01$\textsuperscript{b} & $52.17 \pm 0.4$\textsuperscript{b} & $68.1\pm0.2$\textsuperscript{b,\textdagger}\\
KIC 4671239 & $5295 \pm 145$\textsuperscript{e} & $-2.63 \pm 0.20$\textsuperscript{e} & 0.17\textsuperscript{e} & $-2.51 \pm 0.20$ & $16.78 \pm 2.44$\textsuperscript{f} & $9.82 \pm 0.05$\textsuperscript{g} & $98.9 \pm 1.2$\textsuperscript{g} & $66.67\pm0.06$\textsuperscript{h}\\
KIC 7341231 & $5233 \pm 100$\textsuperscript{i} & $-1.64 \pm 0.10$\textsuperscript{i} & 0.22\textsuperscript{j} & $-1.49 \pm 0.10$ & $5.26 \pm 0.79$\textsuperscript{i} & $28.9 \pm 0.20$\textsuperscript{i} & $406.0 \pm 3.0$\textsuperscript{i} &$111.45\pm0.04$\textsuperscript{*}\\
KIC 8144907 & $5400 \pm 200$\textsuperscript{k} & $-2.66 \pm 0.08$\textsuperscript{k} & 0.38\textsuperscript{k} & $-2.38 \pm 0.08$ & $11.13 \pm 0.45$\textsuperscript{k} & $17.47 \pm 0.09$\textsuperscript{l} & $208.2 \pm 1.1$\textsuperscript{l} & $84.83\pm0.02$\textsuperscript{hk}\\
$\nu$ Indi & $5320 \pm 64$\textsuperscript{m} & $-1.43 \pm 0.09$\textsuperscript{m} & 0.29\textsuperscript{m} & $-1.22 \pm 0.09$ & $6.00 \pm 0.35$\textsuperscript{f} & $25.05 \pm 0.21$\textsuperscript{*} & $350.0 \pm 1.1$\textsuperscript{*} & $101.76 \pm 0.17$\textsuperscript{*}\\
TIC 300085386 & $5211 \pm 55$\textsuperscript{n} & $-1.29 \pm 0.10$\textsuperscript{n}  & 0.32\textsuperscript{n}  & $-1.06 \pm 0.10$ & $20.89 \pm 3.13$\textsuperscript{f} & $6.62 \pm 0.06$\textsuperscript{*} & $60.13 \pm 3.0$\textsuperscript{*} & $66.1\pm0.3$\textsuperscript{*}\\
\hline
\end{tabular}}
\endgroup
\begin{tablenotes}
\footnotesize
\item[Sources:]\textsuperscript{a}\citet{Ishigaki_2012}; \textsuperscript{b}\citet{Lindsay2025}; \textsuperscript{c}\citet{Karovicova2020}; \textsuperscript{d}\citet{Lundkvist2025}; \textsuperscript{e}\citet{Alencastro2022}; \textsuperscript{f}\citet{Gaia_DR2}; \textsuperscript{g}\citet{Larsen2025}; \textsuperscript{h}\citet{Kuszlewicz_2023}; \textsuperscript{i}\citet{Deheuvels2012}; \textsuperscript{j}APOGEE DR16, \citet{apogee}, \citet{APOGEEDR16}; \textsuperscript{k}\citet{Huber2024}; \textsuperscript{l}\citet{Yu_2018}; \textsuperscript{m}\citet{Chaplin2020}; \textsuperscript{n}\citet{deBritoSilva2024}; \textsuperscript{*}This Work; \textsuperscript{\textdagger} assuming RGB phase, see \citet{Lindsay2025}.
\end{tablenotes}
\end{threeparttable}
\end{table*}

Since our goal is to determine how the treatment of $\alpha$-enhancement affects the asteroseismic modeling of evolved metal-poor stars through a homogeneous modeling effort, we build on previous studies that determined the asteroseismic mode frequencies for the target stars and used them to determine the fundamental parameters of the stars. 7 of the 8 stars have been studied using individual mode asteroseismology previously, including HD 128279 \citep{Lindsay2025}, HD 140283 \citep{Lundkvist2025}, HD 175305 \citep{Lindsay2025}, KIC 4671239 \citep{Larsen2025}, KIC 7341231 \citep{Deheuvels2011}, KIC 8144907 \citep{Huber2024}, and $\nu$ Indi \citep{Chaplin2020}. In the case of $\nu$ Indi, additional higher-cadence TESS data has become available since the analysis of \citet{Chaplin2020}, so we determine additional mode frequencies by applying PBJam \citep{PBJam} to all the available data (see \autoref{appendix_nu_indi}). Since TIC 300085386 has not been studied with detailed asteroseismology previously, we determine the oscillation mode frequencies in this work following the same method as described in \citet{Lindsay2025} (see \autoref{appendix_TIC_3000}). In all other cases, we use the individual mode frequencies and errors reported in the previously mentioned corresponding papers in our modeling.

\section{Modeling Methods}
\label{sec:optimization}
We calculate separate sets of evolutionary stellar model tracks for each target star using a similar iterative process to the method described in \citet{Lindsay2025} using the spectroscopic and asteroseismic data described in \autoref{sec:sample} as constraints. We calculate model tracks using MESA version r22.05.1 \citep{Paxton2011,Paxton2013,Paxton2015,Paxton2018,Paxton2019,Jermyn2023}, varying the initial mass ($M_0$), initial helium abundance ($Y_0$), initial metal abundance divided by initial hydrogen abundance ($f = Z_0 / X_0$), and convective mixing length ($\alpha_{\text{mlt}}$) in the following ranges: $0.7 \leq M_0 \leq 1.0$, $0.24 \leq Y_0 \leq 0.28$, $0.00001 \leq f \leq 0.002$, and $1.5 \leq \alpha_{\text{mlt}} \leq 2.5$. The initial parameters of each model track are chosen based on the quality of the fit between the observed properties of each target star and the previously calculated models' spectroscopic and asteroseismic properties. This is done using the differential evolution algorithm implemented in \texttt{yabox} \citep{Yabox}. We use the differential evolution algorithm to choose initial model parameters, as opposed to a grid-based modeling, in order to more densely sample the parameter space around the optimal model parameters. 

We calculate the sets of stellar model tracks for each target star in two different ways, changing only the treatment of $\alpha$ elements. In the $\alpha$-enhanced modeling, we use the same techniques applied in \citet{Lindsay2025}, using element mixtures enhanced in $\alpha$-elements from their GS98 values \citep{GS98} according to the average observed $\alpha$ abundances from \autoref{table:spec_inputs}. The $\alpha$-enhanced models also are computed using correspondingly $\alpha$-enhanced OPAL/Opacity Project opacity tables, included in MESA's kap module. We note that we take the average $\alpha$-enhancement values using the measured elemental abundances of Mg, Si, Ca, and Ti, weighted by the abundance uncertainties \citep[following][]{Lindsay2025, Lundkvist2025} In our $\alpha$-enhanced modeling all $\alpha$-elements are enhanced to the same amount based on the the average [$\alpha$/Fe] value even though in real stars different $\alpha$-elements may be enhanced to different levels compared with the average. For the Salaris-corrected stellar modeling, the MESA model element mixtures are scaled from the GS98 solar values, and the default non-$\alpha$-enhanced GS98 opacity tables are used.

Determining the quality of a model's fit to the observed spectroscopic and asteroseismic data proceeds as follows. We first construct a stellar model evolution track from the pre-main-sequence until the point along the red giant branch where the model's luminosity is significantly higher ($\gtrsim$ 10$\sigma$) than the observed luminosity reported in \autoref{table:spec_inputs}. All models we consider in this work are red giant branch stars, and the Helium burning phases of evolution are not considered. Based on the large frequency separation ($\Delta \nu$) and dipolar period spacing ($\Delta \Pi_{\ell = 1}$) measurements listed in \autoref{table:spec_inputs}, all targets except for HD 175305 and TIC 300085386 have $\Delta \nu$ values too high to be core helium burning stars while HD 175305 and TIC 300085386 have $\Delta \Pi_{\ell = 1}$ values too low to be core helium burning stars. The $\Delta \Pi_{\ell = 1}$ measurements are taken from the literature or, in the case of HD 175305 and TIC 300085386, are made following the methods of \citet{Vrard2016} and \citet{Kuszlewicz_2023}, taking the power spectrum of the stretched-period spectrum. We note that in the case of HD 175305, the period spacing measurement is taken from previous work where we were not able to obtain a fully robust measurement of the dipolar period spacing \citep{Lindsay2025} and are assuming that HD 175305 is a red giant branch star due to its membership in the Helmi Streams and presumed old age. If HD 175305 were a core helium burning star, it would have to be much younger, or be the product of non-standard evolutionary processes. 

In this work, we use MESA's default `basic.net' nuclear reaction network as well as MESA's default settings for the equation of state (EOS) module. The EOS module uses a blend of different equation of state sources (including FreeEOS, OPAL/SCVH EOS, HELM, and SKYE EOS) to determine the thermodynamic properties of the models depending on the conditions of the stellar material \citep[see][]{Jermyn2023}. Model tracks are calculated incorporating diffusion of heavy elements following the prescription of \citet{Thoul1994}, as well as a small amount of extra overshoot mixing beneath the convective envelope following an exponential profile with $f_{\text{ov, exp}} = 0.01 $ and $f_0$ = 0.0005. Envelope overshoot is applied following previous studies that found most models of low-mass red giant branch stars without envelope overshoot fail to reproduce the observed location of the red giant branch luminosity bump \citep{Alongi1991, Khan2018, Lindsay2022}. No extra overshoot mixing was incorporated above convective cores, since our stellar sample contains only low-mass stars which do not have convective cores on the main sequence. 

Along each evolutionary track, we save the global stellar model properties as well as the stellar structure files for models with luminosity, temperature, and [Fe/H] values that agree to within 10$\sigma$ of the values in \autoref{table:spec_inputs}. We determine the cost function output based on the discrepancy between each stellar model and the target star observables, which is calculated as the sum of two $\chi^2$ terms, one spectroscopic term and one asteroseismic term. The spectroscopic term, $\chi^2_{\text{spec}}$, is determined as
\begin{equation}
\label{eq:chi2spec}
\chi^2_{\text{spectroscopic}} = \chi^2_{\text{Luminosity}} + \chi^2_{\text{Temperature}} + \chi^2_{\text{[Fe/H]}}
\end{equation}
with each spectroscopic parameter $P$'s, corresponding $\chi_{P}^2$ value calculated as:
\begin{equation}
\label{eq:chi2}
\chi_{P}^2 = \frac{(P_{\text{obs}} - P_{\text{model}})^2}{\sigma_{P_{\text{obs}}}^2}.
\end{equation}
We find the [Fe/H] values for each $\alpha$-enhanced model by taking the surface metal and hydrogen abundances from each MESA model, then calculating the iron abundance taking into account the different relative abundances of iron in the metal mixture given the $\alpha$-enhancement values reported in \autoref{table:spec_inputs}. For the Salaris-corrected modeling, [Fe/H] is also found from the surface metal and hydrogen abundances, but since each model's metal abundances are solar-scaled, we can take [Fe/H] = [M/H] as there is no $\alpha$-element enhancement. 

In order to determine the fit to the observed asteroseismic data ($\chi^2_{\text{seismic}}$), we calculate the radial, dipole, and quadrupole ($\ell$ = 0, 1, and 2) mode frequencies for models along the evolutionary tracks using the stellar oscillation code GYRE \citep{Townsend2013}. We first calculate the $\ell$ = 0 mode frequencies for all models along the track with $\chi^2_{\text{spectroscopic}} < 100$, then, for models with a close enough match between the observed and model $\ell = 0$ modes ($\chi^2_{\text{seismic}, \ell = 0} < 10$), we also calculate the $\ell = 1$ mixed modes and $\ell = 2$ pure p-modes ($\pi$-modes). The value of $\chi^2_{\text{seismic}, \ell = 0}$ is found by finding the reduced $\chi^2$ between the observed radial mode frequencies and the surface-corrected radial model frequencies using \autoref{eq:chi2seis}. 

We match each model $\ell = 0$ and $\ell = 2$ mode to the observed oscillation modes based on their inferred values of radial order ($n_{\text{p}}$). The observed $n_{\text{p}}$ values are based on the asymptotic relation ($n_{\text{p}} \approx (\nu_{\text{obs}}/\Delta \nu) - (\ell /2)$) and the observed $\Delta \nu$ values from \autoref{table:spec_inputs}, while the model mode $n_{\text{p}}$ values are returned from the GYRE code. As more than one dipolar mixed mode can share the same $n_\text{p}$ value, we match the model $\ell=1$ modes to the observed modes using a nearest neighbor search. The $\ell = 0$ and $\ell = 2$ mode frequencies are corrected for known near-surface modeling errors (the `surface term') using the two-term prescription from \citet{bg14}. The $\ell = 0$ observed and model modes are used to determine the surface-term coefficients, which are then applied to the modes of other angular degrees. The $\ell = 1$ mixed mode frequencies are the result of mode coupling between a core g-mode component, which does not suffer a surface effect, and an envelope p-mode component, which does. Following \citet{ong2020} we apply the surface correction from \citet{bg14} to only the pure-p-mode ($\pi$-mode) component of the mixed-mode frequencies. 

Subtracting the surface-corrected mode frequencies ($\nu_{\text{corr model}, n}$) from the observed frequencies ($\nu_{\text{obs}}$), squaring this difference, and dividing by the squared observed mode frequency errors ($\sigma^2_{\nu_{\text{obs}, n}}$), we find the seismic $\chi^2$ per degree of freedom, $\chi^2_{\text{seismic}}$, for models along the evolutionary track using,
\begin{equation}
\label{eq:chi2seis}
\chi^2_{\text{seismic}} =  \frac{1}{N_\nu - 1 - 2}\sum^{N_{\nu}}_{n} \frac{(\nu_{\text{obs}, n} - \nu_{\text{corr model}, n})^2}{\sigma^2_{\nu_{\text{obs}, n}} },
\end{equation}
where $N_{\nu}$ is the total number of mode frequencies. Since over-corrections in accounting for the surface term can make models whose mode frequencies are far from the observed frequencies seem like good model fits, we add an additional penalty function to $\chi^2_{\text{seismic}}$ calculated from the 2 lowest frequency radial and quadrupole modes. 
\begin{equation}
\label{eq:low_n_cost}
\chi^2_{\text{low }n} = \frac{1}{100} \sum_{\ell\in \{0,2\}} \frac{1}{2} \sum_{n = n_{\text{loweset}}}^{n_{\text{loweset}+1}}\frac{(\nu_{\text{obs}, n} - \nu_{\text{uncorr model}, n})^2}{\sigma^2_{\nu_{\text{obs}, n}} }.
\end{equation}
Similar terms of this type were also included in \citet{BasuKinnane2018}, \citet{Ong2021a}, \citet{Cunha2021}, \citet{Lindsay2024}, and \citet{Lindsay2025} ensuring that the low frequency uncorrected model p-modes agree with the corresponding observed mode frequencies. 

The sum of each model along the track's $\chi^2_{\text{spectroscopic}}$, $\chi^2_{\text{seismic}}$, and $\chi^2_{\text{low }n}$ is the total $\chi^2$ ($\chi^2_{\text{total}}$) and the model with the minimum $\chi^2_{\text{total}}$ is the best-fit model along the evolutionary track. That lowest $\chi^2_{\text{total}}$ is returned to the differential evolution optimization routine and is then used to choose subsequent combinations of initial model parameters in order to more density sample the parameter space in locations where $\chi^2_{\text{total}}$ is low. All $\chi^2_{\text{total}}$ values along each track as well as the model parameters associated model parameters are saved in order to determine the likelihood weighted distributions of the stellar parameters of interest. We calculate the weights of each model based on the $\chi^2_{\text{total}}$ values by taking
\begin{equation}
\label{eq:weight}
W_{\textrm{model}} = \frac{L_{\textrm{total}}}{\sum L_{\textrm{total}}}, \text{ with } L_{\textrm{total}} = \frac{\exp(-\frac{1}{2} \frac{\chi^2_{\text{total}}}{\chi^2_{\text{total, min}}})}{p(M, Y_0, f, \alpha_{\text{mlt}}) / dt}
\end{equation}
where $p(M, Y_0, f, \alpha_{\text{mlt}})$ is the Kernel Density Estimation-based estimate of the local sampling density in mass, initial helium abundance, initial metallicity, and mixing length at that track's values of  $M, Y_0, f = Z_0/X_0, \text{ and } \alpha_{\text{mlt}}$ and $dt$ is the length of each stellar model's timestep. We note that when calculating the likelihoods, we first divide each $\chi^2_{\text{total}}$ value by the minimum $\chi^2_{\text{total}}$ along the entire optimization modeling run for that star. This is done because some stars, particularly those observed by \textit{Kepler}, have very small mode frequency errors, yielding some $\chi^2_{\text{total}}$ that are orders of magnitude too large to calculate an associated $L$ for without running into a floating-point underflow. Tempering all $\chi^2_{\text{total}}$ values by the minimum $\chi^2_{\text{total}}$ value reduces this problem for the \textit{Kepler} stars.

The likelihood weighted parameter distributions are obtained by multiplying the modeling weights ($W_{\textrm{model}}$) by the associated parameters of the model. We report the 16th, 50th, and 84th percentiles of the weighted mass, age, radius, initial helium abundance, convective mixing length, effective temperature, luminosity, and iron abundance distributions determined from the $\alpha$-enhanced and Salaris-corrected modeling for each star in \autoref{table:Results_alpha_enhanced} and \autoref{table:Results_salaris_corrected} respectively. Using the differential evolution algorithm ensures that the region of parameter space around the highest likelihood initial MESA model parameters is more densely sampled than if a grid with uniform sampling was applied, as in most other asteroseismic modeling studies. There is not a minimum $\chi^2_{\text{total}}$ at which the optimization stops. Instead, the modeling stops after 40 iterations of the differential evolution algorithm, corresponding to 840 model evaluations with each evaluation involving the calculation of the MESA evolutionary track, model mode frequencies, and determining the best-fit model and associated $\chi^2_{\text{total}}$. After the modeling finishes, the best fit model among all models from all model tracks is taken as the overall best-fit model. The parameters of the overall best-fit models resulting from the separate $\alpha$-enhanced and Salaris-corrected modeling procedures are reported in \autoref{table:Results_alpha_enhanced_best_fit} and \autoref{table:Results_salaris_corrected_best_fit} respectively.

\begin{figure*}
    \centering
    \includegraphics[width=0.7\textwidth]{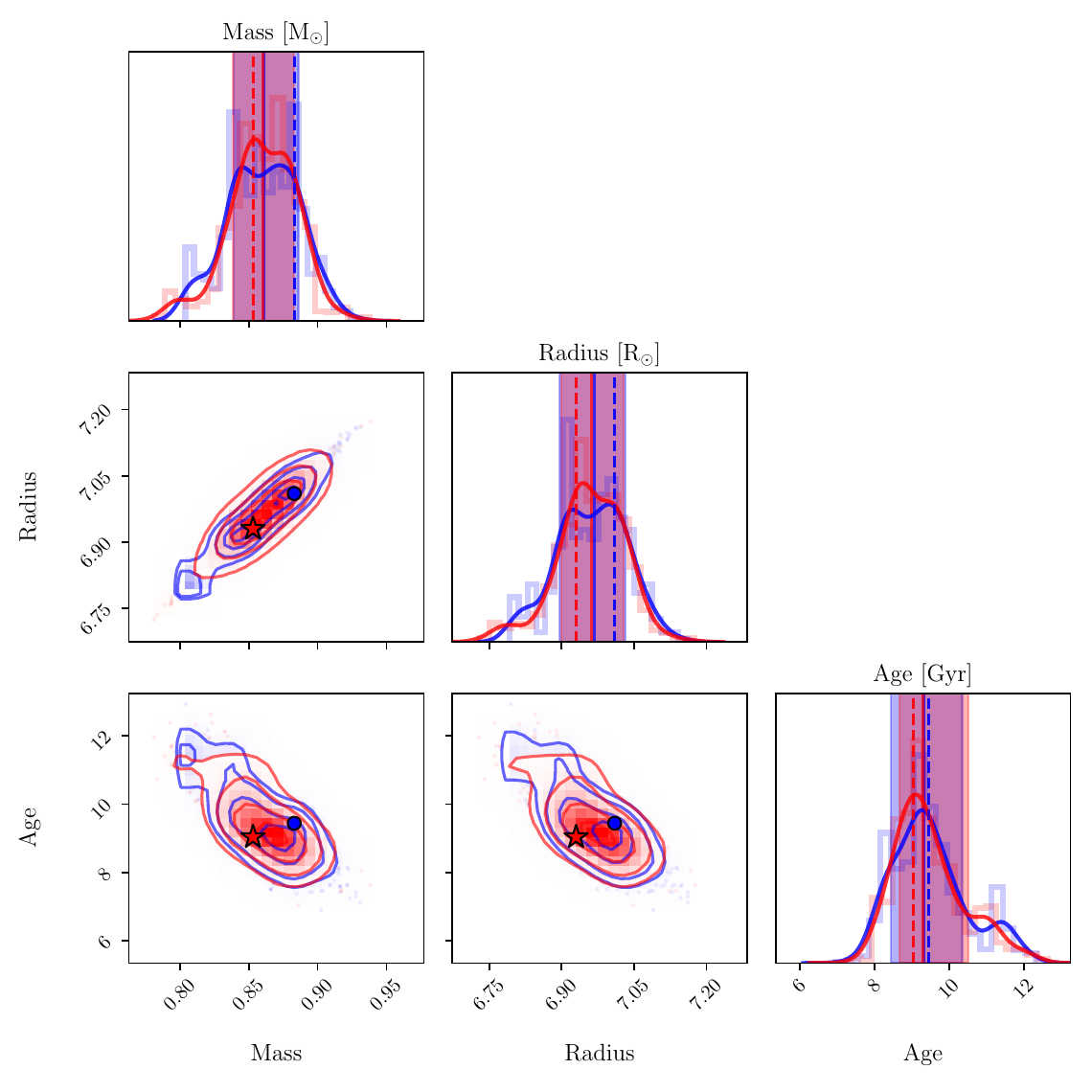}
    \caption{Corner plot visualization showing the mass, radius, and age results of the $\alpha$-enhanced and Salaris-corrected modeling procedure applied to TIC 300085386. The model parameters are taken from the models calculated along each evolutionary track and are weighted by the normalized likelihood (\autoref{eq:weight}) before plotting. The on-diagonal panels show the marginal mass, radius, and age distributions in the form of kernel density estimation (KDE) plots with blue curves showing the $\alpha$-enhanced results and red curves showing the Salaris-corrected results. The vertical blue and red lines over-plotted in the on-diagonal panels show the 50th percentiles of the likelihood-weighted parameter distributions from the $\alpha$-enhanced and Salaris-corrected modeling, respectively, while the blue and red shaded regions span the region between the 16th and 84th percentiles of the same weighted parameter distributions. The vertical blue and red dashed lines show the best-fit $\alpha$-enhanced and Salaris-corrected model parameters, respectively. The blue and red contours in the off-diagonal panels show the joint distributions of different pairs of global stellar parameters derived from the $\alpha$-enhanced and Salaris-corrected modeling procedures, respectively. The data points in the off-diagonal plots show the overall best-fit model parameters from the ompleted $\alpha$-enhanced modeling (blue circle points) and Salaris-corrected modeling (red star points) of TIC 300085386.   } 
    \label{fig:corner_plot_global_params}
\end{figure*}

\begin{figure*}
    \centering
    \includegraphics[width=0.7\textwidth]{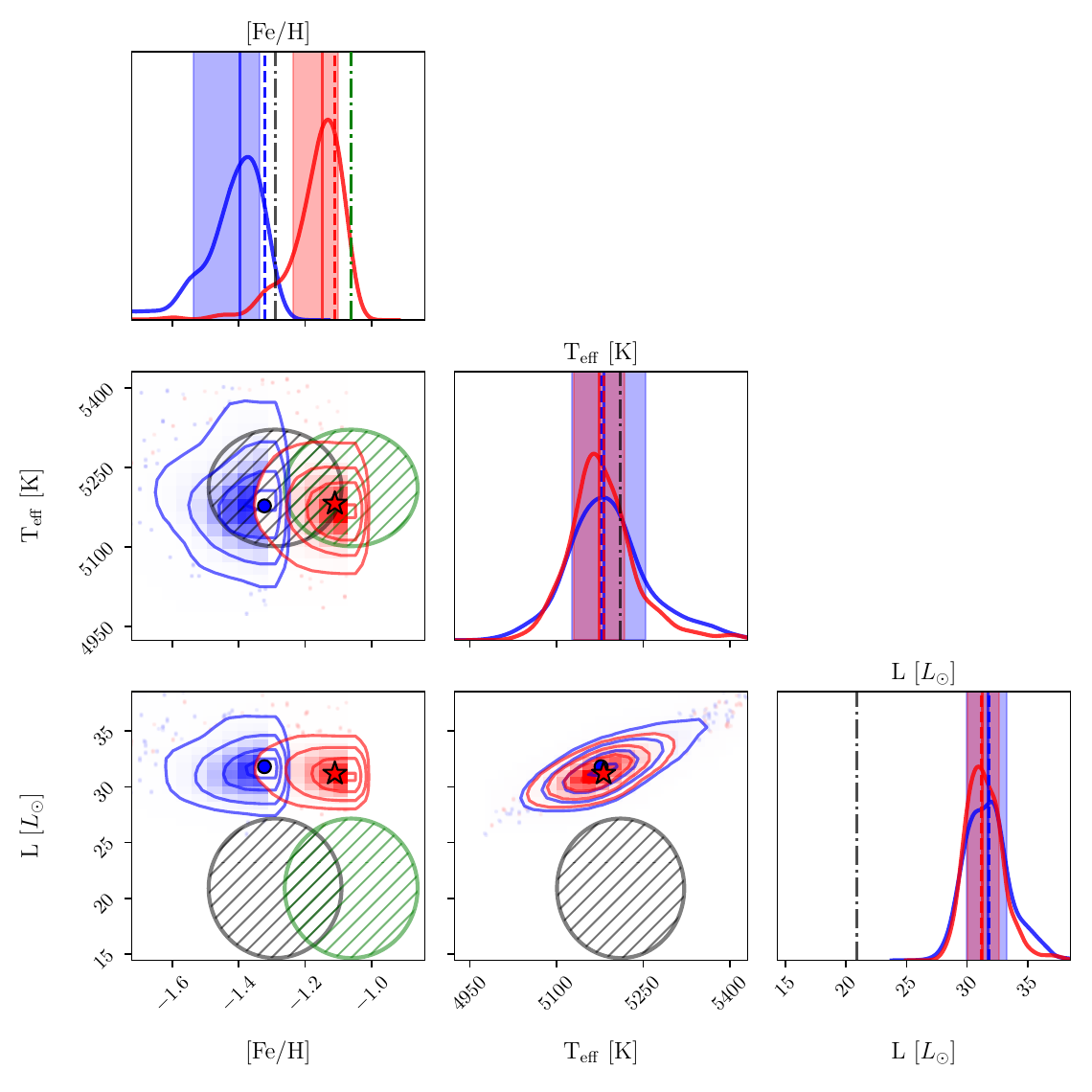}
    \caption{Similar corner plot to \autoref{fig:corner_plot_global_params} except we show the [Fe/H], T$_\text{eff}$, and luminosity likelihood-weighted distributions from the $\alpha$-enhanced and Salaris-corrected modeling procedure applied to TIC 300085386. The vertical black dot-dashed lines in each on-diagonal panel show the observed [Fe/H], T$_\text{eff}$, and luminosity values, with the Salaris-corrected [Fe/H] value shown with a green vertical dot-dashed line. The spectroscopic observations are represented in the off-diagonal joint-distribution panels as 2$\sigma$ gray ellipses. The green 2$\sigma$ ellipses incorporate the Salaris correction to the observed [Fe/H].  } 
    \label{fig:corner_plot_spec}
\end{figure*}

We visualize the resultant weighted parameter distributions from the $\alpha$-enhanced and Salaris-corrected modeling procedure applied to the final target star in our sample, TIC 300085386, in \autoref{fig:corner_plot_global_params} and \autoref{fig:corner_plot_spec}. The corner plots in \autoref{fig:corner_plot_global_params} and \autoref{fig:corner_plot_spec} are constructed by taking the parameters of the models along each track calculated as part of the $\alpha$-enhanced and Salaris-corrected optimization procedures, weighted according to \autoref{eq:weight}. The marginal distribution plots along the diagonal in \autoref{fig:corner_plot_global_params} show the mass, radius, and age distributions in the form of kernel density estimation (KDE) plots for all the $\alpha$-enhanced models in blue and the Salaris-corrected models in red. Similarly, the marginal distribution plots in \autoref{fig:corner_plot_spec} show the [Fe/H], effective temperature, and luminosity distributions for the same $\alpha$-enhanced and Salaris-corrected models. 

In each marginal distribution plot in \autoref{fig:corner_plot_global_params} and \autoref{fig:corner_plot_spec}, the blue and red solid vertical lines show the 50th percentiles of the likelihood-weighted parameter distributions from the $\alpha$-enhanced and Salaris-corrected modeling, respectively, while the blue and red shaded regions span the region between the 16th and 84th percentiles of the same weighted parameter distributions (listed in \autoref{table:Results_alpha_enhanced} and \autoref{table:Results_salaris_corrected}). The vertical dashed blue and red lines in the marginal distribution plots of \autoref{fig:corner_plot_global_params} and \autoref{fig:corner_plot_spec} show the $\alpha$-enhanced and Salaris-corrected best-fit model parameters, respectively. The vertical dot-dashed black lines in \autoref{fig:corner_plot_spec} show the observed values for luminosity, effective temperature, and [Fe/H], with the Salaris-corrected [Fe/H] value shown with a green vertical dot-dashed line. 

The blue and red contours in the off-diagonal panels of \autoref{fig:corner_plot_global_params} and \autoref{fig:corner_plot_spec} show the joint distributions between parameter pairs along with data points showing the overall best-fit model parameters and associated errors from the $\alpha$-enhanced modeling (\autoref{table:Results_alpha_enhanced_best_fit}, blue circle points) and the Salaris-corrected modeling (\autoref{table:Results_salaris_corrected_best_fit}, red star points). The observed spectroscopic parameters listed in \autoref{table:spec_inputs} for TIC 300085386 are represented as gray 2$\sigma$ error ellipses in the joint distribution plots in \autoref{fig:corner_plot_spec}. The green 2$\sigma$ error ellipses show the Salaris-corrected [Fe/H] observations. Equivalent corner plots for the other stars in our sample are shown in \autoref{appendix_corner}.

\begin{table*}[ht!]
\caption{Asteroseismic modeling results for all target stars using the $\alpha$-enhanced modeling method. Each parameter value and range represents the 16th, 50th, and 84th percentiles calculated from the parameters weighted by the normalized likelihood (\autoref{eq:weight}).}
\label{table:Results_alpha_enhanced}
\centering
\begin{tabular}{lccccccccc}
\toprule
Target    & Mass [$\text{M}_{\odot}$] & Age [Gyr]      & Radius [$\text{R}_{\odot}$] & $Y_0$             & $\alpha_{\text{mlt}}$ & T$_{\text{eff}}$ [K] & L [$\text{L}_{\odot}$] & [Fe/H]  & $\Delta \Pi_{\ell = 1}$ [s]         \\
\hline
HD 128279 & $0.766^{+0.041}_{-0.024}$ & $12.9^{+1.2}_{-2.1}$ & $3.732^{+0.079}_{-0.149}$ & $0.258^{+0.013}_{-0.009}$ & $1.86^{+0.26}_{-0.14}$ & $5435^{+78}_{-57}$ & $10.87^{+0.72}_{-0.64}$ & $-2.16^{+0.07}_{-0.07}$ & $\phantom{0}83.74^{+3.07}_{-1.50}$ \\
HD 140283 & $0.751^{+0.038}_{-0.028}$ & $13.4^{+1.7}_{-1.7}$ & $2.036^{+0.070}_{-0.053}$ & $0.259^{+0.010}_{-0.013}$ & $2.13^{+0.18}_{-0.31}$ & $5855^{+89}_{-79}$ & $\phantom{0}4.43^{+0.34}_{-0.40}$ & $-2.16^{+0.37}_{-0.27}$ & $145.17^{+1.86}_{-9.81}$ \\
HD 175305 & $0.791^{+0.036}_{-0.044}$ & $12.3^{+2.8}_{-2.0}$ & $7.291^{+0.130}_{-0.133}$ & $0.261^{+0.009}_{-0.010}$ & $1.99^{+0.18}_{-0.21}$ & $5197^{+94}_{-82}$ & $35.00^{+2.97}_{-2.54}$ & $-1.44^{+0.11}_{-0.26}$ & $\phantom{0}63.66^{+0.82}_{-1.14}$ \\
KIC 4671239 & $0.775^{+0.042}_{-0.027}$ & $12.5^{+1.9}_{-2.4}$ & $5.225^{+0.108}_{-0.070}$ & $0.256^{+0.016}_{-0.010}$ & $1.72^{+0.26}_{-0.14}$ & $5297^{+92}_{-55}$ & $19.57^{+1.56}_{-1.19}$ & $-2.41^{+0.18}_{-0.17}$ & $\phantom{0}69.57^{+1.80}_{-1.73}$ \\
KIC 7341231 & $0.814^{+0.037}_{-0.027}$ & $11.1^{+1.8}_{-1.9}$ & $2.608^{+0.037}_{-0.041}$ & $0.256^{+0.011}_{-0.008}$ & $1.91^{+0.29}_{-0.19}$ & $5505^{+124}_{-104}$ & $\phantom{0}5.60^{+0.77}_{-0.50}$ & $-1.40^{+0.09}_{-0.15}$ & $111.60^{+0.11}_{-0.14}$ \\
KIC 8144907 & $0.775^{+0.015}_{-0.032}$ & $12.7^{+1.3}_{-1.2}$ & $3.582^{+0.021}_{-0.054}$ & $0.255^{+0.010}_{-0.010}$ & $2.09^{+0.17}_{-0.27}$ & $5511^{+58}_{-75}$ & $10.61^{+0.64}_{-0.72}$ & $-2.31^{+0.14}_{-0.31}$ & $\phantom{0}84.82^{+0.07}_{-0.07}$ \\
$\nu$ Indi & $0.800^{+0.011}_{-0.016}$ & $12.9^{+0.6}_{-0.5}$ & $2.812^{+0.018}_{-0.018}$ & $0.248^{+0.005}_{-0.007}$ & $1.76^{+0.04}_{-0.05}$ & $5352^{+21}_{-26}$ & $\phantom{0}5.84^{+0.14}_{-0.14}$ & $-1.41^{+0.03}_{-0.03}$ & $102.96^{+0.18}_{-0.16}$ \\
TIC 300085386 & $0.861^{+0.025}_{-0.022}$ & $\phantom{0}9.3^{+1.0}_{-0.9}$ & $6.967^{+0.066}_{-0.071}$ & $0.258^{+0.011}_{-0.010}$ & $1.87^{+0.18}_{-0.10}$ & $5182^{+71}_{-55}$ & $31.71^{+1.58}_{-1.79}$ & $-1.40^{+0.06}_{-0.14}$ & $\phantom{0}66.61^{+0.70}_{-0.84}$ \\
\hline
\end{tabular}
\end{table*}

\begin{table*}[ht!]
\caption{Asteroseismic modeling results for all target stars using the Salaris-corrected modeling method. Each parameter value and range represents the 16th, 50th, and 84th percentiles calculated from the parameters weighted by the normalized likelihood (\autoref{eq:weight}). }
\label{table:Results_salaris_corrected}
\centering
\begin{tabular}{lccccccccc}
\toprule
Target    & Mass [$\text{M}_{\odot}$] & Age [Gyr]      & Radius [$\text{R}_{\odot}$] & $Y_0$             & $\alpha_{\text{mlt}}$ & T$_{\text{eff}}$ [K] & L [$\text{L}_{\odot}$] & [Fe/H]   & $\Delta \Pi_{\ell = 1}$ [s]          \\
\hline
HD 128279 & $0.761^{+0.023}_{-0.025}$ & $13.4^{+1.2}_{-1.4}$ & $3.739^{+0.064}_{-0.152}$ & $0.258^{+0.013}_{-0.012}$ & $1.89^{+0.14}_{-0.14}$ & $5428^{+55}_{-49}$ & $10.79^{+0.58}_{-0.59}$ & $-1.92^{+0.05}_{-0.06}$ & $\phantom{0}83.45^{+1.83}_{-1.29}$ \\
HD 140283 & $0.750^{+0.033}_{-0.011}$ & $13.0^{+1.7}_{-1.2}$ & $2.076^{+0.028}_{-0.084}$ & $0.262^{+0.015}_{-0.016}$ & $2.06^{+0.19}_{-0.18}$ & $5891^{+66}_{-96}$ & $\phantom{0}4.62^{+0.31}_{-0.44}$ & $-2.00^{+0.22}_{-0.20}$ & $145.59^{+0.71}_{-9.54}$ \\
HD 175305 & $0.797^{+0.037}_{-0.047}$ & $12.1^{+2.7}_{-2.2}$ & $7.311^{+0.120}_{-0.145}$ & $0.261^{+0.012}_{-0.013}$ & $1.81^{+0.32}_{-0.20}$ & $5154^{+124}_{-124}$ & $34.14^{+3.99}_{-4.09}$ & $-1.33^{+0.18}_{-0.24}$ & $\phantom{0}64.00^{+1.30}_{-1.39}$ \\
KIC 4671239 & $0.786^{+0.047}_{-0.038}$ & $11.8^{+2.4}_{-2.4}$ & $5.254^{+0.120}_{-0.097}$ & $0.256^{+0.013}_{-0.009}$ & $1.76^{+0.33}_{-0.18}$ & $5317^{+105}_{-68}$ & $20.05^{+1.86}_{-1.48}$ & $-2.29^{+0.20}_{-0.27}$ & $\phantom{0}69.58^{+1.87}_{-1.76}$ \\
KIC 7341231 & $0.824^{+0.033}_{-0.033}$ & $10.7^{+2.3}_{-1.4}$ & $2.613^{+0.036}_{-0.033}$ & $0.252^{+0.013}_{-0.007}$ & $1.99^{+0.18}_{-0.28}$ & $5538^{+92}_{-152}$ & $\phantom{0}5.76^{+0.51}_{-0.78}$ & $-1.28^{+0.10}_{-0.10}$ & $111.59^{+0.13}_{-0.12}$ \\
KIC 8144907 & $0.786^{+0.024}_{-0.032}$ & $11.8^{+1.5}_{-1.0}$ & $3.598^{+0.040}_{-0.052}$ & $0.254^{+0.015}_{-0.008}$ & $2.21^{+0.11}_{-0.24}$ & $5551^{+51}_{-74}$ & $10.97^{+0.79}_{-0.76}$ & $-2.03^{+0.16}_{-0.32}$ & $\phantom{0}84.82^{+0.09}_{-0.11}$ \\
$\nu$ Indi & $0.793^{+0.016}_{-0.011}$ & $13.0^{+0.4}_{-0.6}$ & $2.808^{+0.017}_{-0.016}$ & $0.249^{+0.008}_{-0.005}$ & $1.76^{+0.05}_{-0.03}$ & $5350^{+26}_{-21}$ & $\phantom{0}5.82^{+0.15}_{-0.11}$ & $-1.21^{+0.04}_{-0.02}$ & $102.99^{+0.18}_{-0.17}$ \\
TIC 300085386 & $0.860^{+0.023}_{-0.022}$ & $\phantom{0}9.3^{+1.2}_{-0.6}$ & $6.961^{+0.067}_{-0.061}$ & $0.261^{+0.013}_{-0.011}$ & $1.88^{+0.10}_{-0.08}$ & $5174^{+45}_{-43}$ & $31.27^{+1.39}_{-1.23}$ & $-1.15^{+0.05}_{-0.09}$ & $\phantom{0}66.73^{+0.52}_{-0.63}$ \\
\hline
\end{tabular}
\end{table*}

\section{Results}
\label{sec:results}

We report the global parameters of the stars in our sample, derived from the likelihood weighted parameter distributions resulting from the $\alpha$-enhanced modeling, in \autoref{table:Results_alpha_enhanced}. The quoted upper and lower uncertainties correspond to the differences between the 84th and 50th percentiles and between the 50th and 16th percentiles, respectively. Similarly, \autoref{table:Results_salaris_corrected} reports the global parameter results for our sample of stars derived using Salaris-corrected models. 

\begin{figure}
    \centering
    \includegraphics[width=0.45\textwidth]{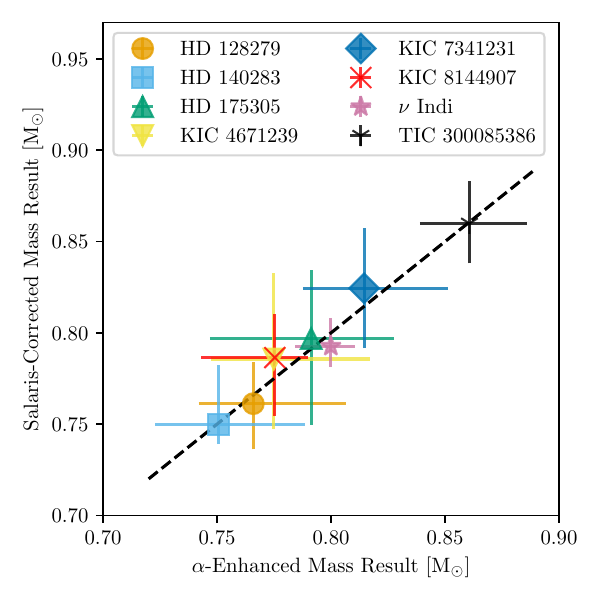}
    \caption{The $\alpha$-enhanced and Salaris-corrected stellar mass modeling results are shown in a `one-to-one' plot. The points are placed at the 50th percentiles of the likelihood-weighted mass distributions while the error bars show the range between the 16th and 84th percentiles listed in \autoref{table:Results_alpha_enhanced} and \autoref{table:Results_salaris_corrected}. The dashed black line shows the line of equality. Both modeling methods produce mass results that agree with each other to within $\sim1\sigma$, indicating that either method for accounting for $\alpha$-enhancement does not significantly change the resultant inferred stellar mass distributions from asteroseismic modeling. } 
    \label{fig:mass_one_to_one}
\end{figure}

\begin{figure}
    \centering
    \includegraphics[width=0.45\textwidth]{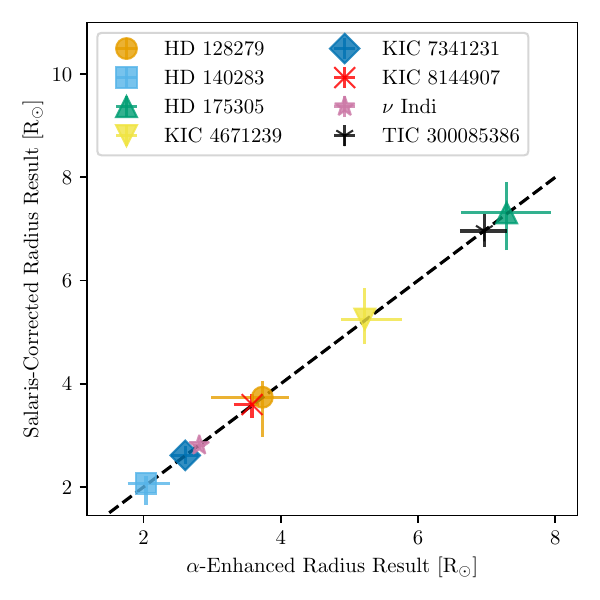}
    \caption{Same as \autoref{fig:mass_one_to_one} except the modeling results for stellar radii are shown, with error bars expanded by 5 times to allow them to be visible. Both modeling methods produce radius results that agree with each other to within $\sim1\sigma$.}
    \label{fig:radius_one_to_one}
\end{figure}

\begin{figure}
    \centering
    \includegraphics[width=0.45\textwidth]{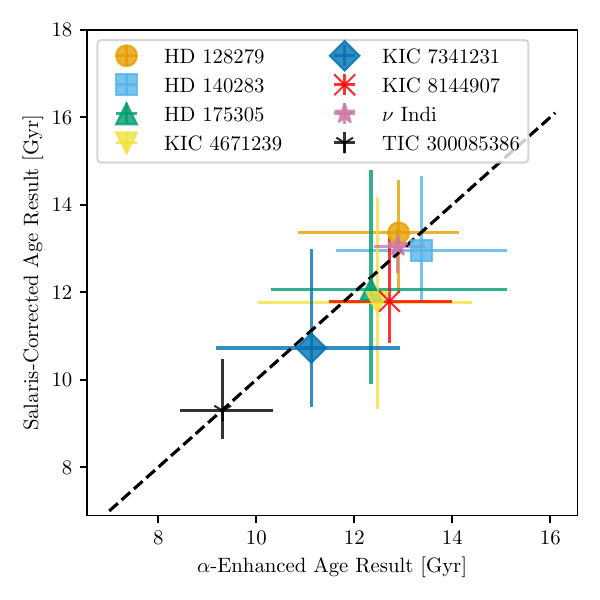}
    \caption{Same as \autoref{fig:mass_one_to_one} except the modeling results for stellar age are shown. Both modeling methods produce stellar age results that agree with each other to within $\sim1\sigma$.} 
    \label{fig:age_one_to_one}
\end{figure}

We find that the global parameters of the stars in our sample obtained using our $\alpha$-enhanced modeling procedure agree well with the parameters obtained by applying the correction of \citet{salaris} to the observed [Fe/H] values and performing the same asteroseismic modeling procedure using non-$\alpha$-enhanced (solar-scaled) stellar models. We illustrate this finding using `one-to-one' plots for the inferred masses (\autoref{fig:mass_one_to_one}), radii (\autoref{fig:radius_one_to_one}), and ages (\autoref{fig:age_one_to_one}). Each `one-to-one' plot displays the mass, radius, or age results derived from the $\alpha$-enhanced modeling on the x-axis and the corresponding mass, radius, or age results from Salaris-corrected modeling on the y-axis, along with the associated errors on the stellar parameters. 

Each star's $\alpha$-enhanced and Salaris-corrected mass results shown in \autoref{fig:mass_one_to_one} agree with each other at the 1$\sigma$ level, meaning all points in the `one-to-one' plot lie along the dashed line of equality when mass errors are taken into account. Similarly, \autoref{fig:radius_one_to_one} shows that all the $\alpha$-enhanced and Salaris-corrected radii results agree to within 1$\sigma$. Finally, the $\alpha$-enhanced and Salaris-corrected age results shown in \autoref{fig:age_one_to_one} show that the age determinations also agree between methods of accounting for $\alpha$-enhancement for all stars in our sample to within 1$\sigma$. Overall, these results show that when using asteroseismic data, both modeling methods for accounting for $\alpha$-enhancement, Salaris-corrected or fully $\alpha$-enhanced, produce very similar mass, radius, and age posterior results. 

The average precision levels on our different stellar parameter results are $\sim$4\% in mass, $\sim$2\% in radius, and $\sim$13\% in age. These levels of precision on mass, radius, and age are generally in line with previous detailed asteroseismic modeling studies of main sequence stars \citep[e.g.][]{Bellinger2019} as well as more evolved stars \citep[e.g.][]{Lindsay2024}. 



\subsection{Best-fit Models}
We also report the global parameters of the best-fit $\alpha$-enhanced models for each target star in \autoref{table:Results_alpha_enhanced_best_fit} and the parameters of the best-fit Salaris-corrected models in \autoref{table:Results_salaris_corrected_best_fit}. The individual best-fit model masses, radii, and ages are visualized with the vertical dashed lines in the on-diagonal panels of the global parameter corner plots (\autoref{fig:corner_plot_global_params} and left-hand plots of the figures in \autoref{appendix_corner}) and with the circle and star points in the off-diagonal joint distribution panels in the global parameter corner plots. The best-fit masses, radii, and ages generally coincide with the dense regions of each parameter's likelihood-weighted distribution, with the vertical dashed lines agreeing with the shaded areas representing the area between the 16th and 84th percentiles of the likelihood-weighted parameter distributions. However, we note that the parameters of the individual best-fit model do not necessarily need to match with the center peaks of the corresponding likelihood weighted parameter distributions since there may be multiple models with different parameters with similarly high likelihoods.

The best-fit model [Fe/H], effective temperature, and luminosity parameters as vertical dashed lines and data points in the spectroscopic quantity corner plots (\autoref{fig:corner_plot_spec} and right-hand plots of the figures in \autoref{appendix_corner}.) As is apparent in \autoref{fig:corner_plot_spec}, the best-fit model's spectroscopic parameters do not always agree exactly with the observed spectroscopic properties. In the case of TIC 300085386, \autoref{fig:corner_plot_spec} shows that the observed luminosity is substantially lower than the best-fit model luminosities obtained from both our $\alpha$-enhanced and Salaris-corrected asteroseismic modeling. Mismatches between the observed spectroscopic parameters and the asteroseismic modeling-based spectroscopic parameters could indicate that the observations are inaccurate; for example, the luminosity of TIC 300085386 is obtained using only GAIA photometry, which is known to give inaccurate luminosities due to a variety of factors, including imperfect dust extinction corrections \citep{Gaia_DR2}. 

The best-fit masses we determine for the stars in our sample are generally low, ranging between 0.74 and 0.9 $M_{\odot}$. Thus, the best-fit ages we infer are hence quite old, with all stars except TIC 300085386 having ages older than 10 Gyr. Since the early enrichment of the Galaxy is thought to be dominated by more core-collapse supernovae, which produce higher concentrations of $\alpha$ elements compared with type 1a supernovae, which occur later. The fact that we infer old ages for our target stars is in line with the fact that all the target stars in our sample are evolved, metal-poor, and enhanced in $\alpha$-elements.  

We note that as in \citet{Lindsay2025}, the age of the universe ($\tau_{\textrm{universe}} \simeq 13.8$ Gyr \citep{Planck2018}) was not explicitly used as a stopping condition in our MESA modeling; however, there is an effective age cutoff enforced through our adopted stellar mass lower bound of $0.7 M_{\odot}$. However, since many stellar models with $M \lesssim 0.77 \text{M}_{\odot}$ are older than $\tau_{\textrm{universe}}$ at the metallicities and evolutionary stages under consideration in this work, it is possible for a target star's best fit model age to be older than the universe. This only occurred for the Salaris-corrected best-fit model for KIC 4671239; however, we find that the age of KIC 4671239 is not in tension with the age of the universe when the width of the weighted distribution of ages is taken into account (\autoref{table:Results_alpha_enhanced} and \autoref{table:Results_salaris_corrected}). This is shown in the left panel of \autoref{fig:corner_plot_global_params_KIC_4671239} which displays that the best-fit Salaris-corrected model age for KIC 4671239 is on the higher age end of the age distribution shown in red, while the 50th percentile of the weighted age distribution is at 12 Gyr, in line with the age of the universe.


\begin{table*}[ht!]
\caption{The global parameters of the individual best-fit $\alpha$-enhanced models for each target star along the optimization trajectory described in \autoref{sec:optimization}. }
\label{table:Results_alpha_enhanced_best_fit}
\centering
\begin{tabular}{lcccccccc}
\toprule
Target    & Mass [$\text{M}_{\odot}$] & Age [Gyr]      & Radius [$\text{R}_{\odot}$] & $Y_0$             & $\alpha_{\text{mlt}}$ & T$_{\text{eff}}$ [K] & Luminosity [$\text{L}_{\odot}$] & [Fe/H]           \\
\hline
HD 128279  &  0.768 & 13.7 & 3.799 & 0.246 & 1.766 & 5374 & 10.85 & -2.21 \\
HD 140283  &  0.776 & 12.8 & 2.109 & 0.248 & 1.911 & 5846 & \phantom{0}4.68 & -2.28 \\
HD 175305  &  0.794 & 11.7 & 7.309 & 0.267 & 1.947 & 5193 & 35.00 & -1.43 \\
KIC 4671239  &  0.781 & 11.2 & 5.271 & 0.269 & 1.507 & 5226 & 18.67 & -2.39 \\
KIC 7341231  &  0.777 & 13.8 & 2.565 & 0.242 & 1.655 & 5418 & \phantom{0}5.11 & -1.68 \\
KIC 8144907  &  0.79 & 11.5 & 3.603 & 0.257 & 2.184 & 5569 & 11.25 & -2.62 \\
$\nu$ Indi  &  0.815 & 12.5 & 2.828 & 0.242 & 1.778 & 5356 & \phantom{0}5.93 & -1.41 \\
TIC 300085386  &  0.883 & \phantom{0}9.4 & 7.01 & 0.248 & 1.944 & 5177 & 31.81 & -1.32 \\
\hline
\end{tabular}
\end{table*}

\begin{table*}[ht!]
\caption{The global parameters of the individual best-fit Salaris-corrected models for each target star along the optimization trajectory described in \autoref{sec:optimization}. }
\label{table:Results_salaris_corrected_best_fit}
\centering
\begin{tabular}{lcccccccc}
\toprule
Target    & Mass [$\text{M}_{\odot}$] & Age [Gyr]      & Radius [$\text{R}_{\odot}$] & $Y_0$             & $\alpha_{\text{mlt}}$ & T$_{\text{eff}}$ [K] & Luminosity [$\text{L}_{\odot}$] & [Fe/H]           \\
\hline
HD 128279  &  0.776 & 13.4 & 3.815 & 0.244 & 1.786 & 5380 & 10.98 & -1.98 \\
HD 140283  &  0.740 & 13.0 & 2.074 & 0.273 & 1.872 & 5840 & \phantom{0}4.51 & -2.02 \\
HD 175305  &  0.846 & 11.0 & 7.470 & 0.241 & 1.940 & 5176 & 36.08 & -1.27 \\
KIC 4671239  &  0.754 & 14.2 & 5.188 & 0.251 & 1.749 & 5298 & 19.11 & -2.43 \\
KIC 7341231  &  0.791 & 12.5 & 2.579 & 0.249 & 1.826 & 5496 & \phantom{0}5.47 & -1.49 \\
KIC 8144907  &  0.810 & 10.8 & 3.636 & 0.252 & 2.284 & 5606 & 11.77 & -2.48 \\
$\nu$ Indi  &  0.797 & 12.9 & 2.808 & 0.248 & 1.777 & 5362 & \phantom{0}5.87 & -1.22 \\
TIC 300085386  &  0.853 & \phantom{0}9.0 & 6.930 & 0.274 & 1.897 & 5182 & 31.19 & -1.11 \\
\hline
\end{tabular}
\end{table*}

\subsection{Dipolar Gravity Mode Period Spacings}
A red giant star's asymptotic gravity mode period spacing ($\Delta\Pi_{\ell}$) refers to the spacing between consecutive g-modes of high radial order ($n$) and the same spherical degree, $\ell$. $\Delta\Pi_{\ell}$ is sensitive to the size and buoyancy frequency profile of the stellar core and is calculated from stellar models as  
\begin{equation}
    \label{eq:period_spacing}
    \Delta \Pi_{\ell} = \frac{2 \pi^2}{\sqrt{\ell (\ell + 1)}} \left( \int_{\textrm{core}} \frac{N}{r}dr\right)^{-1}
\end{equation}
where $\ell$ is the angular degree, $r$ is the radius coordinate, $N$ is the Brunt–Väisälä frequency, and the integral is taken over the extent of the radiative core of a star \citep{Tassoul1980}. At the same value of $\Delta \nu$ (10 $\mu$Hz), we found that the model $\Delta \Pi_{\ell = 1}$ values do change, depending on the model treatment of $\alpha$-enhancement, with $\Delta \Pi_{\ell = 1} = 79.2$ s for the $\alpha$-enhanced model and $\Delta \Pi_{\ell = 1} = 78.8$ s for the Salaris-corrected model (see \autoref{fig:alpha_methods}). This difference is considerably larger than measurement errors on $\Delta\Pi_1$ known to be achievable from fitting the parameters of the asymptotic mixed-mode eigenvalue relation \citep[e.g.][]{Mosser_2018,Kuszlewicz_2023,ong_mode_2023,li_rotation_2024} against \textit{Kepler} data, although comparable to that attainable from e.g. a few sectors of \textit{TESS} data.

To test if applying different treatments of $\alpha$-enhancement changes the inferred model asymptotic dipolar period spacing, we applied \autoref{eq:period_spacing} to the models we calculate in our modeling procedure to determine posterior distributions of $\Delta\Pi_{\ell = 1}$ resulting from the $\alpha$-enhanced and Salaris-corrected modeling of the stars in our sample. We report the posterior predictive $\Delta\Pi_{\ell = 1}$ results for our $\alpha$-enhanced and Salaris-corrected modeling in the last columns of \autoref{table:Results_alpha_enhanced} and \autoref{table:Results_salaris_corrected} respectively. Overall, we find that the $\Delta\Pi_{\ell = 1}$ results from both modeling methods agree closely (within the 1$\sigma$ level for all stars in the sample), meaning the $\alpha$-enhanced and Salaris-corrected stellar models, which match the observed stellar properties, have similar core properties as well. 

However, we note that the (posterior predictive) 1$\sigma$ errors on $\Delta\Pi_{\ell = 1}$ we derive from our modeling are all larger than the typical measurement uncertainties on $\Delta\Pi_{\ell = 1}$ obtained from analysis of \textit{Kepler} data from fitting parameterized expressions rather than constructing stellar models --- in some cases considerably so. Observed errors on $\Delta\Pi_{\ell = 1}$ can be very small, given the high quality of data available. For example, \citet{Huber2024} and \citet{Kuszlewicz_2023} found $\Delta\Pi_{\ell = 1} = 84.83 \pm 0.02$ s by modeling the power spectrum of KIC 8144907. In our $\alpha$-enhanced and Salaris-corrected modeling of KIC 8144907, we find $\Delta\Pi_{\ell = 1} = 84.69^{+0.22}_{-0.22}$ s and $\Delta\Pi_{\ell = 1} = 84.79^{+0.26}_{-0.19}$ s, respectively: within 1$\sigma$ agreement with the $\Delta\Pi_{\ell = 1}$ measurement from \citet{Huber2024}, but only with respect to our much larger posterior uncertainties. 

\section{Discussion}
\label{sec:discussion}

\subsection{Asteroseismic Scaling Relations versus Individual Frequency Modeling}

Our results qualitatively disagree with those of \cite{morales_model_tayar}, who find that stellar ages inferred from 1D stellar-evolution models depend strongly on modeling assumptions --- including, in particular, opacity tables and chemical mixtures, both of which are varied in this work. One possible reason for discrepancy is that, while asteroseismic constraints in \cite{morales_model_tayar} were provided only by $\Delta\nu$ and $\numax$, and then only through scaling relations, those in this work did not rely on scaling relations at all --- our models were constrained instead only by numerical frequencies of oscillation, and without $\numax$. We note that our age uncertainties are larger than reported in that work, despite our seismic constraints on internal structure (and therefore model age) being stronger.

Previous works that used individual mode frequencies to determine the global parameters of metal-poor stars have found that the results obtained by using only the global asteroseismic parameters ($\Delta \nu$ and $\nu_{\text{max}}$) disagree significantly with results obtained using individual mode frequencies \citep[e.g.][]{Huber2024, Larsen2025, Lindsay2025, Lundkvist2025}. In particular, scaling relation-based masses for metal-poor stars are significantly higher than masses obtained using detailed frequency modeling. Other works focused on applying the global asteroseismic scaling relations to stars in the Milky Way's Halo have also found that global asteroseismology gives stellar mass results that are larger than expected for old, metal-poor Halo stars \citep[e.g.][]{Epstein2014}, highlighting the need for detailed asteroseismology when analyzing metal-poor stars.  

Recent work examining the ramifications of surface effects on the asteroseismic scaling relation for $\Delta \nu$ \citep[$\Delta \nu \propto \sqrt{\rho}$,][]{Ulrich1986} has found that the systematic offsets in $\Delta \nu$ resulting from near-surface modeling errors are not able to fully account for the large variations between the results from global seismology and individual frequency modeling \citep{LiTanda2022, LiYaguang_2023, Huber2024}. This indicates that the scaling relation for $\nu_{\text{max}}$ \citep[$\nu_{\text{max}} \propto g \text{T}_{\text{eff}}^{-1/2}$,][]{Brown1991, Kjeldsen1995} fails at low metallicities. In this work, we use the same method to determine the global parameters for 8 metal-poor targets with individual mode frequencies, permitting a homogeneous determination of how the measured $\nu_{\text{max}}$ values differ from the $\nu_{\text{max}}$ values implied by the inferred mass, radius, and effective temperature results listed in \autoref{table:Results_alpha_enhanced} and \autoref{table:Results_salaris_corrected}. We quantify this deviation by dividing the observed $\nu_{\text{max}}$ value by the predicted $\nu_{\text{max}}$ value that one would obtain by inserting the model mass, radius, and temperature values into the $\nu_{\text{max}}$ scaling relation, \cref{eq:numax_scaling} \citep[see also][]{sharma, LiYaguang2024_realistic, Huber2024, Larsen2025, Lindsay2025}. Thus, $f_{\nu_{\text{max}}}$ is calculated as
\begin{equation}\label{eq:f_nu_max}
    f_{\nu_{\text{max}}} = \frac{\nu_{\text{max}}}{\nu_{\text{max}, \odot}} \bigg{(}\frac{M}{\text{M}_{\odot}}\bigg{)}^{-1} \bigg{(}\frac{R}{\text{R}_{\odot}}\bigg{)}^{2} \bigg{(}\frac{T_{\text{eff}}}{\text{T}_{\text{eff},\odot}}\bigg{)}^{1/2}.
\end{equation}

\begin{figure*}[htbp]
    \centering
    \includegraphics[width=0.48\textwidth]{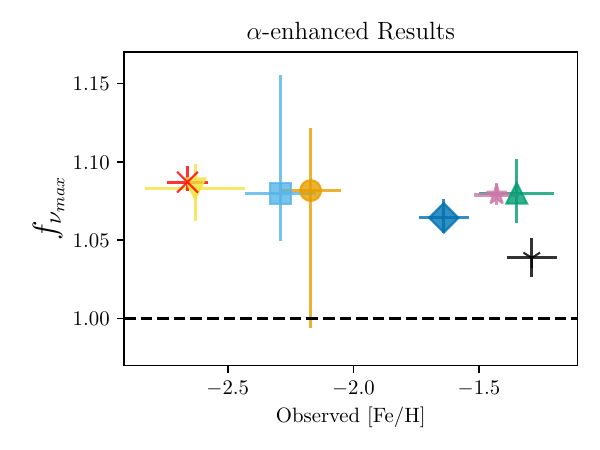}
    \hspace{0.01\textwidth}
    \includegraphics[width=0.48\textwidth]{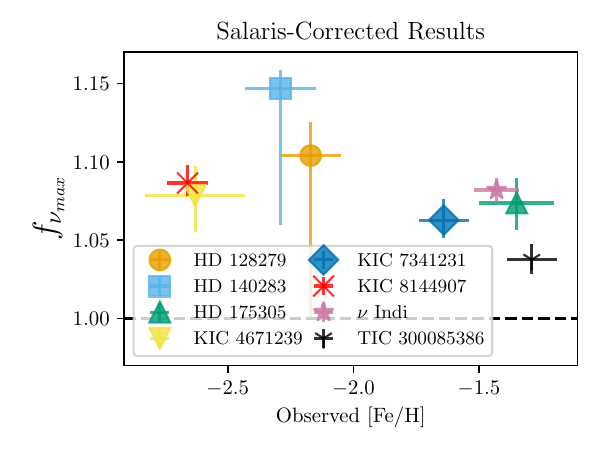}
    \caption{The $f_{\nu_{\text{max}}}$ results inferred using our $\alpha$-enhanced modeling results (left) and Salaris-corrected results (right). The error bars represent the modeling [Fe/H] errors on the x-axis and the 16th to 84th percentiles calculated from the likelihood-weighted $f_{\nu_{\text{max}}}$ distributions for each star.}
    \label{fig:f_nu_max_results}
\end{figure*}

Using $\nu_{\text{max}, \odot} = 3090\mu$Hz, $\text{T}_{\text{eff},\odot} = 5778$ we plug in the mass, radius, and temperature of each model along every evolutionary track calculated as part of the $\alpha$-enhanced or Salaris-corrected modeling procedure. Then we determine the likelihood weighted distribution of $f_{\nu_{\text{max}}}$ in the same way we find the weighted distributions of the other parameters, determining the 16th, 50th, and 84th percentiles. These results are listed in \autoref{table:f_nu_max_results} and we visualize the $f_{\nu_{\text{max}}}$ percentile results for our sample of stars in \autoref{fig:f_nu_max_results}, with each panel showing the $f_{\nu_{\text{max}}}$ results as a function of the observed [Fe/H]. The left and right panels of \autoref{fig:f_nu_max_results} show the $f_{\nu_{\text{max}}}$ results derived from our $\alpha$-enhanced and Salaris-corrected modeling procedure, respectively. Since our mass, radius, and temperature results derived from using $\alpha$-enhanced and Salaris-corrected stellar models are largely similar, the panels of \autoref{fig:f_nu_max_results} are largely the same; however, the errors in $f_{\nu_{\text{max}}}$ are generally larger in the Salaris-corrected due to the larger parameter errors obtained in our Salaris-corrected modeling. 

\begin{table}[ht!]
\caption{The $f_{\nu_{\text{max}}}$ values and ranges for all target stars using the $\alpha$-enhanced and Salaris-corrected modeling methods. The values and ranges come from the 16th, 50th, and 84th percentiles of the likelihood-weighted $f_{\nu_{\text{max}}}$ distributions. }
\label{table:f_nu_max_results}
\centering
\begin{tabular}{lcccccccc}
\toprule
Target    &  $f_{\nu_{\text{max}}}$ ($\alpha$-enhanced) & $f_{\nu_{\text{max}}}$ (Salaris corrected) \\
\hline
HD 128279  &  $1.08^{+0.04}_{-0.09}$  &  $1.10^{+0.02}_{-0.10}$ \\
HD 140283  &  $1.08^{+0.08}_{-0.03}$  &  $1.15^{+0.01}_{-0.09}$ \\
HD 175305  &  $1.08^{+0.02}_{-0.02}$  &  $1.07^{+0.02}_{-0.02}$ \\
KIC 4671239  &  $1.08^{+0.02}_{-0.02}$  &  $1.08^{+0.02}_{-0.02}$ \\
KIC 7341231  &  $1.06^{+0.01}_{-0.01}$  &  $1.06^{+0.01}_{-0.01}$ \\
KIC 8144907  &  $1.09^{+0.01}_{-0.01}$  &  $1.09^{+0.01}_{-0.01}$ \\
$\nu$ Indi  &  $1.08^{+0.01}_{-0.01}$  &  $1.08^{+0.01}_{-0.01}$ \\
TIC 300085386  &  $1.04^{+0.01}_{-0.01}$  &  $1.04^{+0.01}_{-0.01}$ \\
\hline
\end{tabular}
\end{table}

The results shown in \autoref{fig:f_nu_max_results} show that our modeling corroborates the findings of \citet{Huber2024}, \citet{Larsen2025}, and \citet{Lundkvist2025}. These previous studies all found for the metal-poor stars they modeled, that the observed $\nu_{\text{max}}$ values were larger than the $\nu_{\text{max}}$ values obtained by imputing the masses, radii, and temperatures they derived using individual frequency modeling into the $\nu_{\text{max}}$ scaling relation, resulting in similar positive values of $f_{\nu_{\text{max}}}$. 

\subsection{Comparison with Original Modeling Works for the Stars in our Sample}

The goal of this work was to perform a homogeneous modeling of 8 metal-poor stars with two different methods of accounting for enhancement in $\alpha$-elements. We performed our own modeling for each of the stars. However, since 7 of the 8 stars were previously studied using detailed asteroseismology, we compare our results with those previous works in this section and broadly find that despite any differences in modeling methods, our results agree with the various previous asteroseismic analyses of these stars. 

\subsubsection{HD 175305 and HD 128279}

Although the modeling procedure used to determine the parameters of HD 175305 and HD 128279 in \citet{Lindsay2025} was largely the same as the $\alpha$-enhanced modeling method carried out in \autoref{sec:optimization}, we expanded the permitted ranges for some of the input stellar model parameters in this work. Also, since in this work we were interested in how the inferred likelihood weighted stellar parameter distributions change with different treatments of $\alpha$-enhancement, we report the medians and $\pm 1\sigma$ uncertainties of the likelihood-weighted parameter distributions as well as the parameters of the overall best-fit model parameters. In \citet{Lindsay2025}, only the parameters of the best fit model are reported. Overall, the best-fit model parameters from \citet{Lindsay2025} for HD 128279 and HD 175305 agree well with the asteroseismic modeling results from both the $\alpha$-enhanced and Salaris-corrected modeling methods employed in this work (listed in \autoref{table:Results_alpha_enhanced}) and \autoref{table:Results_salaris_corrected} respectively.

For HD 128279, we find that the mass ($0.766^{+0.041}_{-0.024} M_{\odot}$) and radius ($3.732^{+0.079}_{-0.149} R_{\odot}$) results we determined using the $\alpha$-enhanced modeling method agree to 1$\sigma$ with the results from \citet{Lindsay2025} ($M = 0.77 \pm 0.02 M_{\odot}$, $R = 3.80 \pm 0.03 R_{\odot}$). The corresponding age determined in \citet{Lindsay2025} (12.5 $\pm$ 1 Gyr) is also consistent with the age we determine using $\alpha$-enhanced models in this work ($12.9^{+1.2}_{-2.1}$ Gyr), with agreement at the 1$\sigma$ level. There is similar agreement between the Salaris-corrected modeling results (\autoref{table:Results_salaris_corrected}) and the HD 128279 results from \citet{Lindsay2025}.

For HD 175305, we find that the mass ($0.791^{+0.036}_{-0.044}$ $M_{\odot}$) and radius ($7.291^{+0.130}_{-0.133}$ $R_{\odot}$) we determine in this work using the $\alpha$-enhanced modeling method in this work agree to 1$\sigma$ with the HD 175305 results from \citet{Lindsay2025} ($M = 0.83 \pm 0.02 M_{\odot}$, $R = 7.4 \pm 0.07R_{\odot}$). The ages determined for HD 175305 in both \citet{Lindsay2025} ($11.16 \pm 0.91$ Gyr) and this work ($12.3^{+2.8}_{-2.0}$ Gyr) are also in 1$\sigma$ agreement. 


\subsubsection{HD 140283}

In \citet{Lundkvist2025}, individual frequency modeling of HD 140283, also known as the Methuselah star, was performed using a dense grid of models calculated with the stellar evolution code GARSTEC \citep{GARSTEC}. As in our modeling, the authors vary the model initial mass, initial helium abundance, initial metallicity, and convective mixing length, though they also allow the $\alpha$-element enhancement to vary while we keep the $\alpha$-enhancement fixed to a single value in both our $\alpha$-enhanced and Salaris-corrected modeling methods. Using the effective temperature, iron abundance, parallax, and individual radial and dipole mode frequencies as constrains, a mass of $0.75 \pm 0.01 M_{\odot}$, a radius of $2.078^{+0.012}_{-0.011} R_{\odot}$, and an age of 14.2 $\pm$ 0.4 Gyr was determined for HD 140283 in \citet{Lundkvist2025}. They also find $Y_0 = 0.254^{+0.004}_{-0.005}$ and $\alpha_{\text{mlt}} = 1.80^{+0.05}_{-0.06}$. Based on the results of \citet{Lundkvist2025}, applying the same formula for $f_{\nu_{\text{max}}}$ as we do (\autoref{eq:f_nu_max}), a value of $1.14 \pm 0.03$ was determined, which is in agreement with the $f_{\nu_{\text{max}}}$ we determine for HD 140283 ($1.08^{+0.08}_{-0.03}$  and  $1.15^{+0.01}_{-0.09}$ from our $\alpha$-enhanced and Salaris-corrected modeling, respectively). 

Overall there is good agreement between the BASTA \citep{BASTA} modeling carried out in \citet{Lundkvist2025} and this work, which is expected since, although we use a MESA/GYRE-based approach, the input effective temperature, iron abundance, and mode frequencies are the same. The mass, age, and radius values we determine for HD 140283 using both the $\alpha$-enhanced and Salaris-corrected modeling methods agree within 1$\sigma$ with the results from \citet{Lundkvist2025} (see \autoref{table:Results_alpha_enhanced} and \autoref{table:Results_salaris_corrected}). We note that although we find $M = 0.75M_{\odot}$ for HD 140283 in both our $\alpha$-enhanced and Salaris-corrected modeling, in agreement with \citet{Lundkvist2025}, our age results tend to be lower as we determine the age of HD 140283 to be about 13 Gyr. One explanation for this could be that in our modeling, the convective mixing length parameter was allowed to vary up to $\alpha_{\text{mlt}} = 2.5$ whereas the upper bound for $\alpha_{\text{mlt}}$ in the grid based modeling carried out in \citet{Lundkvist2025} was 1.90. We find $\alpha_{\text{mlt}} \approx 2.1$ based on our modeling which is higher than that maximum.

\subsubsection{KIC 4671239}
The fundamental parameters of KIC 4671239 were determined in \citet{Larsen2025} using a BASTA-based modeling procedure. They find a mass of $0.78^{+0.04}_{-0.03}$, a radius of $5.26^{+0.09}_{-0.07}$, and an age of $12.1^{+1.6}_{-1.5}$ Gyr for KIC 4671239. There is close ($<1\sigma$) agreement between the BASTA modeling results from \citet{Larsen2025} and the results we determine in this work using both the $\alpha$-enhanced and Salaris-corrected MESA/GYRE-based modeling (listed in \autoref{table:Results_alpha_enhanced} and \autoref{table:Results_salaris_corrected}), though we note that the errors in stellar mass, radius, and particularly age we determine are larger than those determined in \citet{Larsen2025}. \citet{Larsen2025} determine $f_{\nu_{\text{max}}}$ = 1.101, which agrees to 1$\sigma$ with our determination of $f_{\nu_{\text{max}}}$ from our $\alpha$-enhanced ($f_{\nu_{\text{max}}}$ = $1.08^{+0.02}_{-0.02}$) and Salaris-corrected ($f_{\nu_{\text{max}}} = 1.08^{+0.02}_{-0.02}$) modeling.

\subsubsection{KIC 7341231}

Since the determination of $\Delta\Pi_1$ in \citet{Deheuvels2012} for KIC 7341231 predates the development of more recent techniques for estimating it from observational data \citep[e.g.][]{Mosser_2018,REGGAE,PBJam2.0,liagre_near_2025}, we re-fit its value from the full 4-year \textit{Kepler} short-cadence power spectrum, using the same techniques as we apply to $\nu$ Indi (described in \autoref{appendix_nu_indi}), and finding that $\Delta\Pi_1 = 111.45 \pm 0.04\ \mathrm{s}$. Our statistical uncertainties are consistent with that obtained by \cite{liagre_near_2025}, although our reported value is slightly higher, which is possibly a result of differences in how we parameterize the pure g-mode frequencies at low radial order (as we include a curvature term in the asymptotic relation). Our updated measurement is also more consistent with the values emerging from our stellar models.

\citet{Deheuvels2012} calculated best-fit models for KIC 7341231 at varying metallicities ranging from [Z/X] = -1.75 to [Z/X] = -0.75 using the stellar evolution code \texttt{cesam2k} \citep{Morel1997}. Based on the assumed metallicity, they determined the mass of KIC 7341231 to be in a range between 0.77 and 0.88 $M_{\odot}$, the radius to be between 2.55 and 2.67 $R_{\odot}$, and the age range to be between 11.3 and 14.3 Gyr. Accounting for our uncertainties, the mass, radius, and age results of both our $\alpha$-enhanced and Salaris-corrected listed in \autoref{table:Results_alpha_enhanced} and \autoref{table:Results_salaris_corrected} are within the mass, radius, and age ranges determined in \citet{Deheuvels2012} though we note that our age estimates are on the younger side of the age range determined in \citet{Deheuvels2012}, and that their models were constructed to possess different values of $\Delta\Pi_1$ from those returned from our grid).

\subsubsection{KIC 8144907}
Asteroseismic modeling of KIC 8144907 was carried out in \citet{Huber2024} using a variety of stellar evolution codes and modeling methods, including MESA/GYRE. Overall, they find a mass of $0.79 \pm 0.02 \text{ (ran)} \pm 0.01\text{ (sys)}$, a radius of $3.62 \pm 0.04 \text{ (ran)} \pm 0.02\text{ (sys)}$, and an age of $12.0 \pm 0.6 \text{ (ran)} \pm 0.4\text{ (sys)}$ Gyr for KIC 8144907 and also calculate an $f_{\nu_{\text{max}}}$ value of $1.056 \pm 0.03$ based on their best-fit model. Both the stellar mass and radius we find using $\alpha$-enhanced modeling ($M = 0.775^{+0.015}_{-0.032} M_{\odot}$, $R = 3.582^{+0.021}_{-0.054} R_{\odot}$) are slightly lower than the adopted mass and radius found for KIC 8144907 based on the BeSPP-derived best-fit GARSTEC \citep{GARSTEC} model detailed in \citet{Huber2024}. Consequently, the age we derive from the $\alpha$-enhanced modeling of KIC 8144907 ($\tau = 12.7^{+1.3}_{-1.2}$ Gyr) is slightly older than the adopted age value from \citet{Huber2024}. Overall, though, the masses, radii, and ages agree between the works at the $1\sigma$ level. 

There is also agreement between our Salaris-corrected results for KIC 8144907 and the adopted KIC 8144907 parameters from \citet{Huber2024}. The Salaris-correction based modeling procedure produces a mass ($0.786^{+0.024}_{-0.032} M_{\odot}$), radius ($3.598^{+0.040}_{-0.052} R_{\odot}$), and age ($11.8^{+1.5}_{-1.0}$ Gyr) that agree with the results from \citet{Huber2024} to within 1$\sigma$.

\subsubsection{$\nu$ Indi}
\citet{Chaplin2020} performed detailed asteroseismic modeling of the halo star $\nu$ Indi using just 1 sector of 2-minute cadence data from TESS, yielding 18 mode frequencies. Like \citet{Huber2024}, multiple methods involving different modeling and oscillation codes were used to find the fundamental parameters of the target star. \citet{Chaplin2020} reports a mass of $M = 0.85 \pm 0.04 \text{ (stat)} \pm 0.02\text{ (sys)} M_{\odot}$ and an age of $\tau = 11 \pm 0.7 \text{ (stat)} \pm 0.8\text{ (sys)}$ for $\nu$ Indi. No radius determination for $\nu$ Indi was reported, nor measurement of $\Delta\Pi_1$. 

Since $\nu$ Indi was observed for more TESS sectors in a faster (20-second) cadence, we used the 20-second data to determine 49 oscillation mode frequencies, more than twice the number of modes compared with \citet{Chaplin2020} (see \autoref{appendix_nu_indi}). Overall, after including the additional asteroseismic data, the masses and ages we determine for $\nu$ Indi using our $\alpha$-enhanced and Salaris-corrected modeling methods still agree to within 1$\sigma$ errors compared with the mass and age determined for $\nu$ Indi in \citet{Chaplin2020}. We find mass results that are slightly smaller than the adopted masses from \citet{Chaplin2020} ($\sim$0.80 $M_{\odot}$ versus 0.85 $M_{\odot}$), and consequently we find a slightly higher age compared with the age adopted in \citet{Chaplin2020} ($\sim$13 Gyr versus 11 Gyr).

Both of our model grids produce posterior estimates for the period spacing which differ significantly from the model-independent measured value. This implies that our best-fitting models possess internal structures which differ from those of the true star. Given that the inferred age is a property of the stellar structure, this may in turn also suggest an underlying undiagnosed systematic offset in the inferred age, not necessarily associated with model treatment of $\alpha$-enhancement. While this might be worrying at first, given that the asteroseismic age of $\nu$ Indi has in turn previously been used in \cite{Chaplin2020} to age-date the stellar stream with which it is kinematically associated, the effect of this is still small relative to the (large) uncertainties on its age.


\subsection{Period spacings and mixed modes in stellar modeling}

In addition to being valuable for determining the evolutionary state and histories of red giant stars \citep[e.g.][]{Bedding2011,Rui2021}, precise measurements of a star's dipolar gravity mode period spacing ($\Delta\Pi_{\ell = 1}$) determined from the observed mixed-mode spectrum provides an additional constraint on the structure of the star's interior, improving the precision of asteroseismic modeling results. We have found that our individual-frequency modeling produces reported posterior uncertainties that are an order of magnitude larger than what is otherwise observationally possible to determine without the use of stellar models.

The large $\Delta\Pi_{\ell = 1}$ uncertainties we report arise due to incompleteness in the observed set of dipolar mixed modes, in combination with the use of a nearest-neighbor matching scheme to compare the observed and model dipolar mode frequencies. Since only the most p-mode-dominated dipolar mixed-mode frequencies are observable, we are not able to match every model mixed mode to a corresponding observed mode. This means potentially erroneous matches between mixed modes of different $g$-mode radial order ($n_g$) --- such as might occur where one more or one less model mixed mode exists in intervals where the dipole modes of the true star are not observable --- may still result in low $\chi^2_{\text{seismic}}$ values being computed from nearest-neighbor frequency matching. Equivalently, our posterior distributions are mixtures of much narrower conditional ones under different scenarios of mode identification for the g-modes \citep[cf. section 4.2 and fig. 14 of][]{ong2020}. Therefore, models with significantly different values of $\Delta\Pi_{\ell = 1}$ from the true star may still have mode frequencies within the observable ranges that closely match the observed ones under nearest-neighbor matching. Similar concerns were reported by \cite{campante_rg_2023} for more luminous red giants than considered here. This is of particular concern in the case of HD 140283, where the TESS data yield only a very small number of dipolar mixed modes. Thus, the posteriors we obtain for $\Delta\Pi_{\ell = 1}$ are wide since they are made from models where the computed mixed modes may be of entirely different identities than those in the true star but are assigned high likelihoods all the same.

Ultimately, this indicates that the apparent robustness of our results may stem from a limitation of frequency-modeling techniques that currently are in common use, such as those we have employed, rather than because stellar structures truly are robust against $\alpha$-enhancement. Indeed, the measurably large differences in $\Delta\Pi_1$ between our models from different treatments (as shown in \autoref{fig:alpha_methods}) suggests measurable differences in structure, which are simply not reflected in the uncertainties produced by our modeling pipeline. Conversely, this suggests that if a frequency-modeling technique were to be devised which were to alleviate these limitations --- e.g. by incorporating $\Delta\Pi_1$ directly into the modeling constraints, and/or preserving g-mode identification --- the modeling results from such an improved procedure would be able to discriminate between different treatments of $\alpha$-enhancement, and therefore that an accurate treatments of $\alpha$-enhancement would be necessary for use with such improved techniques.

%


\section{Summary and Conclusion}
\label{sec:Summary}

In this work, we have conducted a homogeneous asteroseismic modeling of 8 metal-poor stars ([Fe/H] $\lesssim 1.5$) that are enhanced in $\alpha$-elements (0.15 $\lesssim$ [$\alpha$/Fe] $\lesssim$ 0.4). Our objective was to test two different methods of accounting for $\alpha$-enhancement and determine if incorporating an $\alpha$-element-enhanced element mixture with correspondingly $\alpha$-enhanced opacity tables would result in different derived stellar parameters compared to modeling stars using the widely used global `Salaris-correction' \citep{salaris} to the observed metallicity.

We performed our asteroseismic modeling of the 8 stars (HD 128279, HD 140283, HD 175305, KIC 4671239, KIC 7341231, KIC 8144907, $\nu$ Indi, and TIC 300085386) using the two aforementioned ways of accounting for the $\alpha$-enhancement when creating stellar models using MESA. In our detailed asteroseismic modeling, we use spectroscopic ([Fe/H], $T_{\text{eff}}$, and luminosity measurements) as well as individual mode frequencies from various sources in the literature for all targets in our sample except $\nu$ Indi and TIC 300085386 (for which mode frequencies were fitted anew for this work). 


From our modeling, we find:

\begin{enumerate}
    \item Comparing the detailed asteroseismic modeling results from our $\alpha$-enhanced and Salaris-corrected modeling procedures shows that the two different methods of accounting for $\alpha$-element enhancement in stellar modeling codes result in very similar derived stellar masses, radii, and ages. The uncertainties on the derived stellar parameters are also similar between the modeling methods.
    \item Our modeling results corroborate recent studies that found that the stellar masses obtained by using only the global asteroseismic parameters are overestimated when compared to masses obtained using detailed asteroseismic modeling \citep{Huber2024, Larsen2025, Lindsay2025}. Using the results from our homogeneous modeling, we confirm that for metal-poor stars, the $\nu_{\text{max}}$ values measured directly from the oscillation power spectra are consistently higher than the $\nu_{\text{max}}$ values implied by applying the $\nu_{\text{max}}$ scaling relation to the masses, radii, and temperatures obtained from detailed asteroseismic modeling.
    \item The close agreement between our modeling results and the stellar parameters derived for the same stars in our sample using different modeling methods shows broadly that asteroseismic modeling produces consistent results at low metallicities even when different modeling codes and methods are used. 
\end{enumerate}

However, we have also found that the modeling procedure we have used produces posterior uncertainties on the g-mode period spacing, and thus the sizes of radiative cores in red giants, that are an order of magnitude larger than observational uncertainties on it. This suggests our procedure to be less sensitive to the structure of these interior parts of stars than what is possible with the data in hand. We have also seen that different treatments of $\alpha$-enhancement produce values of the period spacing that are discrepant by differences more than an order of magnitude larger than these measurement uncertainties, despite being smaller than our reported posterior ones. Therefore, constraints on internal stellar structure may \textit{not} be robust to different treatments of $\alpha$-enhancement, and the above results may also not necessarily apply with improved methods of incorporating structural constraints from p/g mixed modes into stellar modeling.

These findings are in qualitative disagreement with recent findings that red-giant stellar ages derived from 1D stellar evolution depend strongly on choices of model opacities and abundances \citep{morales_model_tayar}. However, our results agree with the work carried out in \citet{apok2} which found that using the correction from \citet{salaris} produces stellar ages which agree with ages determined using stellar models calculated with $\alpha$-enhanced opacities. Our modeling procedure differs from these works chiefly in the use of individual pulsation-mode frequencies, rather than quantities returned from scaling relations, to constrain the internal structure (rather than global properties) of these red giants. This contributes to the mounting body of evidence that the use of scaling relations alone may not suffice for asteroseismic constraints on stellar properties, in particular ages.

Future asteroseismic studies will produce high-quality light curves for an ever-increasing population of stars across the different Milky Way substructures. With the advent of the Roman Space Telescope \citep{Roman_Asteroseismology, WeissDowning2025} and the PLAnetary Transits and Oscillations of stars mission \citep[PLATO][]{Plato_mission}, the asteroseismic study of metal-poor, $\alpha$-enhanced stars across the Galaxy will benefit from an influx of new data. The asteroseismology of metal-poor, $\alpha$-enhanced stars will be crucial for unraveling the formation history of the Galaxy’s oldest components, such as the stellar halo and thick disk.

\begin{acknowledgements}
We thank the anonymous referee, whose comments yielded substantial improvements to our paper. We would like to thank M. Lundkvist and J. Larsen for providing the asteroseismic and spectroscopic properties of HD 140283. We would also like to thank A. Stokholm and D. Huber for their helpful discussion. CJL acknowledges support from a Gruber Science Fellowship. CJL and SB acknowledge support from NSF grant AST-2205026. JMJO acknowledges support from NASA through the NASA Hubble Fellowship grant HST-HF2-51517.001, awarded by the Space Telescope Science Institute (STScI). S.G. acknowledges support by the National Aeronautics and Space Administration under grants 80NSSC23K0137 and 80NSSC23K0168 issued through the TESS Guest Investigator Program. M.H. acknowledges support from NASA grant 80NSSC24K0228. We acknowledge the use of public TESS data from pipelines at the TESS Science Office and at the TESS Science Processing Operations Center. The TESS data presented in this article were obtained from the Mikulski Archive for Space Telescopes (MAST) at the Space Telescope Science Institute. The specific observations analyzed can be accessed via \dataset[doi: 10.17909/pmmr-6z79]{https://doi.org/10.17909/pmmr-6z79}.

This research has made use of the Exoplanet Follow-up Observation Program website, which is operated by the California Institute of Technology under contract with the National Aeronautics and Space Administration under the Exoplanet Exploration Program. Funding for the TESS mission is provided by NASA’s Science Mission Directorate.

\software{
MESA \citep{Paxton2011,Paxton2013,Paxton2015,Paxton2018,Paxton2019,Jermyn2023}, GYRE \citep{Townsend2013}},
SciPy \citep{scipy}, Pandas \citep{pandas}, Astropy \citep{Astropy1, Astropy2, Astropy3}, Lightkurve \citep{Lightkurve}, \pbjam \citep{PBJam, PBJam2.0}, \reggae \citep{REGGAE}, TACO \citep{taco}, YABOX \citep{Yabox}, corner.py \citep{corner}

Example MESA and GYRE inlists we used to generate our models and calculate model frequencies are archived on Zenodo and can be downloaded at \dataset[https://zenodo.org/records/16995052]{https://zenodo.org/records/16995052}.
We also include the MESA evolutionary tracks and profiles for our best-fit models.

\end{acknowledgements}

\bibliographystyle{aasjournalv7}
\bibliography{biblio}

\appendix 

\section{Individual Mode Frequency Determination for $\nu$ Indi}

Previous asteroseismic studies of $\nu$-Indi used only 27 days (one sector) of TESS data  \citep{Chaplin2020}, taken at an observational cadence of 120 s. Since then, 5 additional 27-day-long sectors of TESS observations have been taken of $\nu$-Indi. For two of these sectors, sector 67 and sector 68, $\nu$-Indi was observed in 20-second cadence. 20-second cadence data allows for improved precision compared to 120 second data \citep{Huber2022}. 

\begin{figure}
    \centering
    \includegraphics[width=0.7\textwidth]{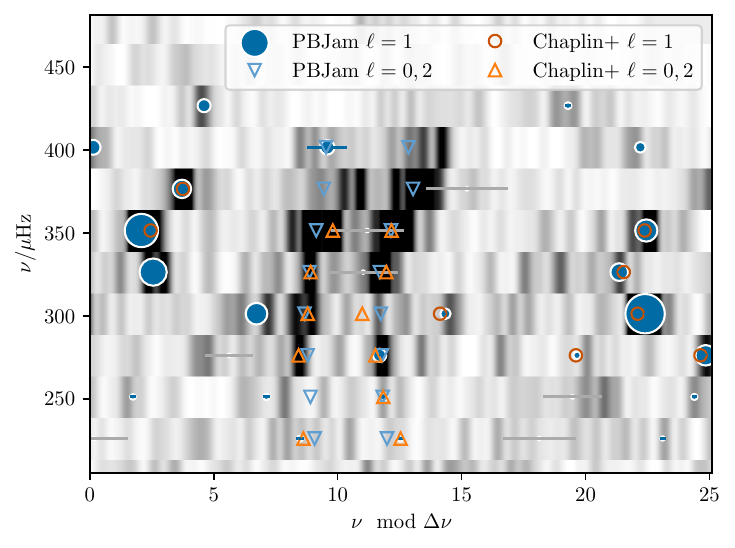}
    \caption{Échelle diagram (frequency versus frequency mod large frequency separation) showing modes fitted against the 20-second cadence $\nu$ Indi TESS data. The radial and quadrupole p-mode frequencies we fitted using \texttt{PBJam} are shown with blue triangles and the dipolar mixed-mode frequencies we fitted using \texttt{reggae} are shown with blue circles. The size of the blue circle represents the fitted mixed-mode amplitude. Grey intervals show where \texttt{reggae} predicts dipole modes which either are too low in amplitude, or lie too close to an even-degree mode, to permit unambiguous identification and constraints from the power spectrum. The background heat-map shows the power spectrum of $\nu$ Indi, and the orange markers show the $\ell = 0, 1$ and 2 modes identified in \citet{Chaplin2020} using sine-wave fitting in the time domain.}
    \label{fig:mode_fit}
\end{figure}

In order to determine the individual mode frequencies from the TESS timeseries, we first stitched the 20-second cadence data from sectors 67 and 68 together using the \texttt{Lightkurve} package \citep{Lightkurve}. Then, after calculating the power spectra of the stitched lightcurve, the radial and quadrupole ($\ell = 0, 2$) p-mode frequencies were calculated using the open-source peak-bagging package \texttt{PBJam} \citep{PBJam}. Mode identifications for the radial and quadrupole frequencies were validated using a different peak-bagging code, \texttt{TACO} \citep{taco}. The dipolar ($\ell = 1$) mixed mode frequencies were identified and fitted using \texttt{reggae}, which implements a generative model for the $\ell = 1$ mixed modes using the $\pi$- and $\gamma$-mode parameterization described in \citet{ong2020}. The g-mode period spacing $\Delta\Pi_1$ is fitted as part of this mode identification procedure, and we list the resulting value in \cref{table:spec_inputs}. The p-mode and mixed-mode frequencies are listed in \autoref{table:nu_indi_modes} and shown in \autoref{fig:mode_fit} along with the previously identified modes from \citet{Chaplin2020}.

\label{appendix_nu_indi}

\section{Individual Mode Frequency Determination for TIC 300085386}
\label{appendix_TIC_3000}

The global asteroseismic parameters and individual oscillation mode frequencies were determined from the TESS observations using the peak-bagging code TACO \citep{taco}. TIC 300085386 was observed for numerous sectors due to its position in the southern continuous viewing zone. We used the \texttt{giants} pipeline \citep{Giants_Pipeline} to produce noise-corrected TESS light curves following the de-trending procedure described in Section 2.2 of \citet{Saunders2022} and smoothing procedure described in Section 3.3 of \citet{Grunblatt2021}. The \texttt{giants} pipeline was designed for the detection of transiting planet \citep{Grunblatt2022,Grunblatt2023, Grunblatt2024,Pereira2024,Saunders2024Planet,Saunders2025} and asteroseismic \citep{Grunblatt2021, Grunblatt2022, Saunders2025} signals in the light curves of evolved stars,  Then, the global asteroseismic parameters, $\Delta \nu$ and $\nu_{\text{max}}$ were calculated using TACO by first calculating a power density spectrum (PDS) from the light curve, dividing out the background components arising from convective granulation and white noise, then finding the frequency of maximum oscillation power ($\nu_{\text{max}}$). Oscillation peaks are then identified from the power density spectrum by applying a Mexican-hat wavelet-transform based algorithm to iteratively find resolved peaks. The peaks are then fitted with Lorentzian functions to determine the frequencies and frequency uncertainties. 

The pressure mode oscillation modes ($\ell = 0$ and $\ell = 2$) are identified using the universal pattern for p-mode oscillations of the same angular degree, $\nu_{n_p, \ell, m} \sim \Delta \nu \big(n_p + \frac{\ell}{2} + \epsilon_p\big)$, where $\Delta \nu$ is the large frequency spacing, $\epsilon_p$ is a phase term, and $n_p$, $\ell$, and $m$ are the p-mode radial order, angular degree, and azimuthal order respectively \citep{Tassoul1980}. The large frequency spacing is identified by TACO using the spacing in frequency between $\ell=0$ p-modes of consecutive radial order. For the peaks not identified using the universal pattern as $\ell = 0$ or $\ell = 2$ modes, TACO identifies them as dipole modes ($\ell = 1$). The frequencies, frequency errors, and angular degrees for the oscillation modes of TIC 300085386 are reported in \autoref{table:TIC_300085386_modes}.

\begin{table}[h!]
\centering
\begin{minipage}{0.48\textwidth}
\centering
\begin{tabular}{lccc}
\toprule
$\ell$                   & $\nu$ {[}$\mu$Hz{]}            & $\sigma_{\nu}$ {[}$\mu$Hz{]}  \\
\hline
0      & 237.970             & 0.745                                                         \\
0      & 262.907             & 0.736                                                          \\
0      & 287.988             & 0.658                                                       \\
0      & 313.046             & 0.646                                                         \\
0      & 338.120             & 0.075                                                         \\
0      & 363.679             & 0.192                                                       \\
0      & 389.679             & 0.386                                                       \\
0      & 414.600             & 0.711                                                       \\
1      & 234.490             & 0.146                                                        \\
1      & 238.509             & 0.099                                                          \\
1      & 249.096             & 0.148                                                         \\
1      & 252.821             & 0.134                                                         \\
1      & 258.203             & 0.167                                                        \\
1      & 262.951             & 0.064                                                         \\
1      & 275.465             & 0.105                                                         \\
1      & 287.892             & 0.073                                                        \\
1      & 295.871             & 0.089                                                         \\
1      & 301.048             & 0.072                                                         \\
1      & 308.016             & 0.313                                                        \\
1      & 315.652             & 0.099                                                        \\
1      & 323.716             & 0.044                                                        \\
1      & 328.979             & 0.059                                                        \\
1      & 347.780             & 0.063                                                       \\
1      & 353.605             & 0.049                                                        \\
1      & 373.979             & 0.060                                                        \\
1      & 380.353             & 0.056                                                         \\
1      & 401.870             & 0.123                                                         \\
1      & 411.280             & 0.803                                                        \\
1      & 423.953             & 0.100                                                         \\
1      & 431.473             & 0.085                                                       \\
1      & 446.132             & 0.145                                                         \\
2      & 235.050             & 0.768                                                        \\
2      & 259.990             & 0.755                                                        \\
2      & 284.989             & 0.696                                                       \\
2      & 309.971             & 0.360                                                          \\
2      & 335.279             & 0.222                                                         \\
2      & 360.661             & 0.565                                                        \\
2      & 386.077             & 0.729                                                        \\
2      & 411.276             & 0.758                                                        \\
\hline 
\end{tabular}
\caption{$\nu$-Indi mode frequencies extracted from the two sectors of 20-second cadence TESS data.\label{table:nu_indi_modes}}
\end{minipage}\hfill
\begin{minipage}{0.48\textwidth}
\centering
\begin{tabular}{lccc}
\toprule
$\ell$                   & $\nu$ {[}$\mu$Hz{]}            & $\sigma_{\nu}$ {[}$\mu$Hz{]}  \\
\hline
0&	46.832&	0.007\\
0&	53.237&	0.015\\
0&	59.809&	0.078\\
0&	66.485&	0.007\\
0&	73.276&	0.007\\
1&	50.006&	0.120\\
1&	50.127&	0.045\\
1&	50.260&	0.059\\
1&	56.143&	0.210\\
1&	56.352&	0.055\\
1&	56.525&	0.056\\
1&	56.719&	0.108\\
1&	56.890&	0.100\\
1&	62.663&	0.059\\
1&	62.897&	0.070\\
1&	63.128&	0.296\\
1&	63.332&	0.057\\
1&	63.561&	0.070\\
1&	69.812&	0.065\\
1&	70.118&	0.099\\
1&	76.647&	0.199\\
1&	76.868&	0.207\\
2&	45.881&	0.008\\
2&	52.212&	0.016\\
2&	58.889&	0.015\\
2&	65.409&	0.021\\
\hline 
\end{tabular}
\caption{TIC 300085386 mode frequencies extracted from 30-minute cadence TESS data.\label{table:TIC_300085386_modes} }
\end{minipage}
\end{table}

\clearpage
\section{Corner Plots}
\label{appendix_corner}
\begin{figure*}
    \centering
    \includegraphics[width=0.49\textwidth]{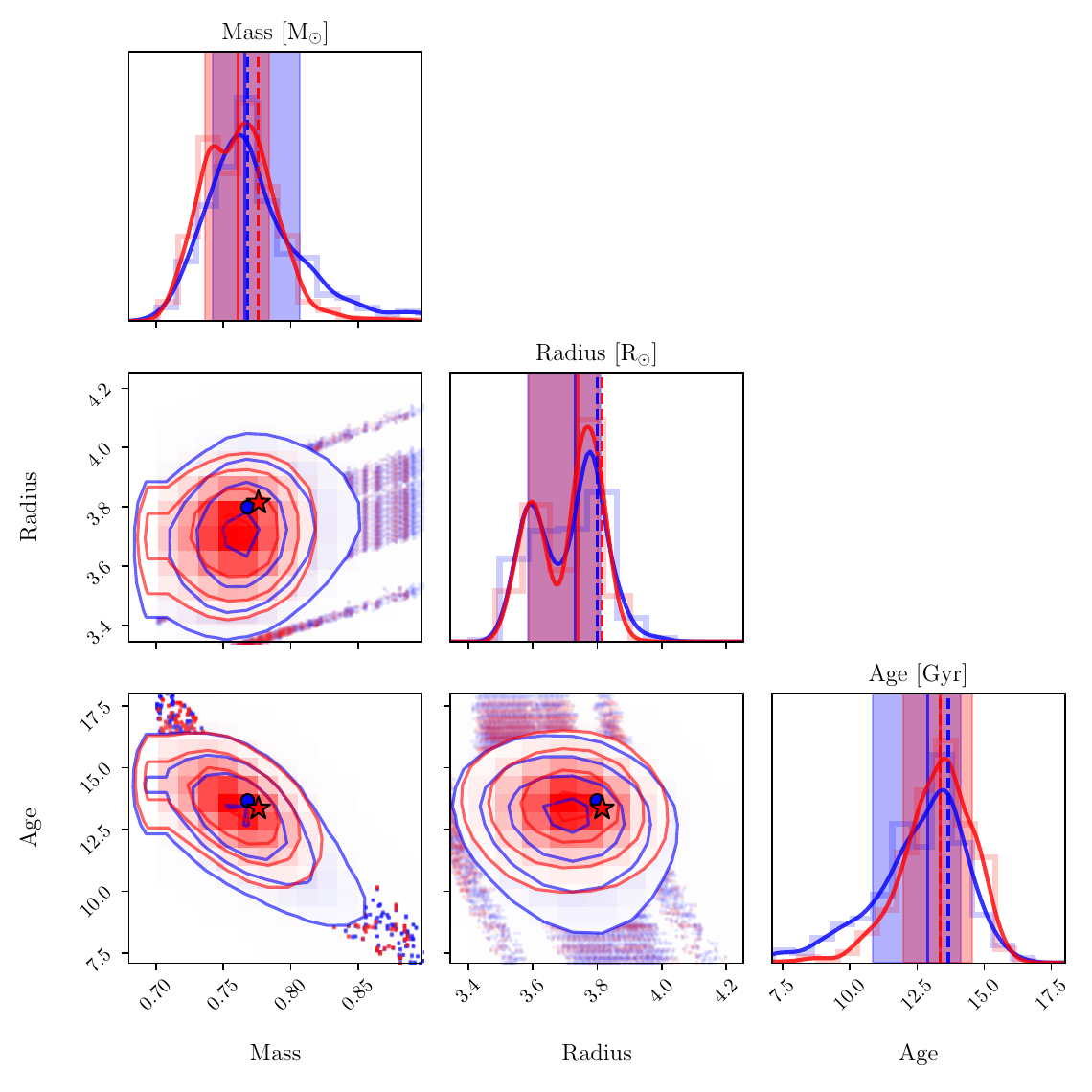}
    \includegraphics[width=0.49\textwidth]{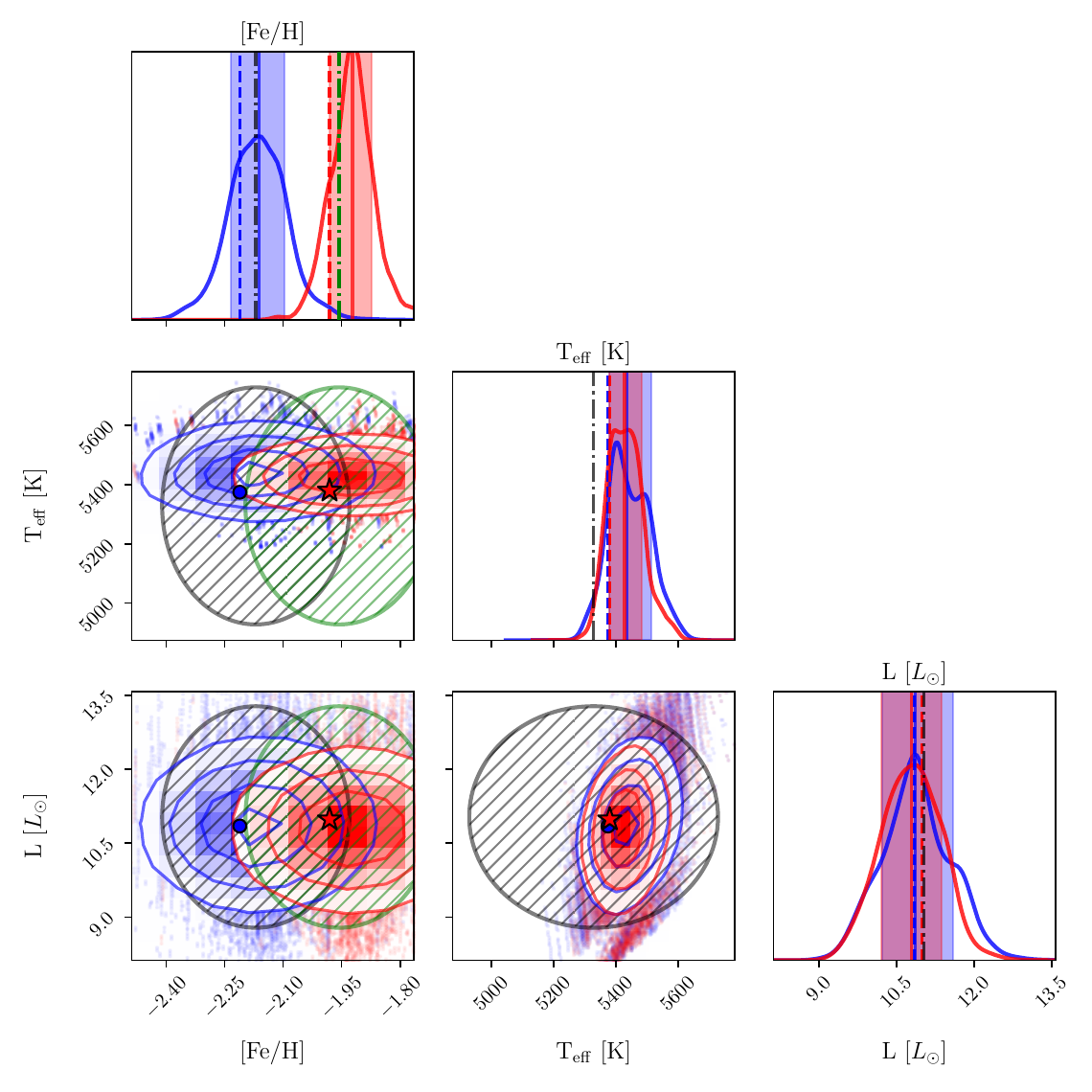}
    \caption{Left panel shows the global parameter corner plot similar to \autoref{fig:corner_plot_global_params} except for HD 128279. The right panel shows the spectroscopic parameter corner plot similar to \autoref{fig:corner_plot_spec} except for HD 128279. } 
    \label{fig:corner_plot_global_params_HD_128279}
\end{figure*}

\begin{figure*}
    \centering
    \includegraphics[width=0.49\textwidth]{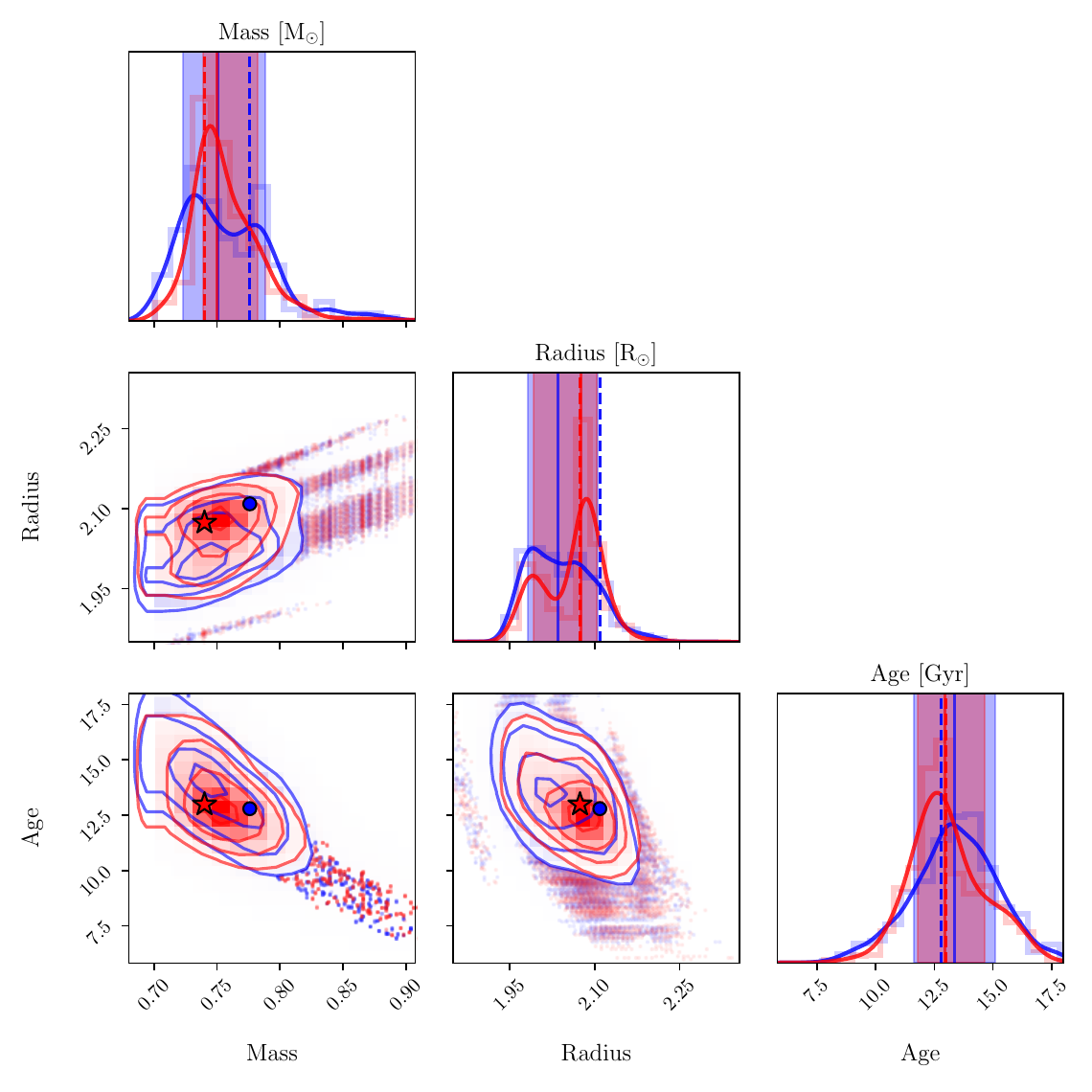}
    \includegraphics[width=0.49\textwidth]{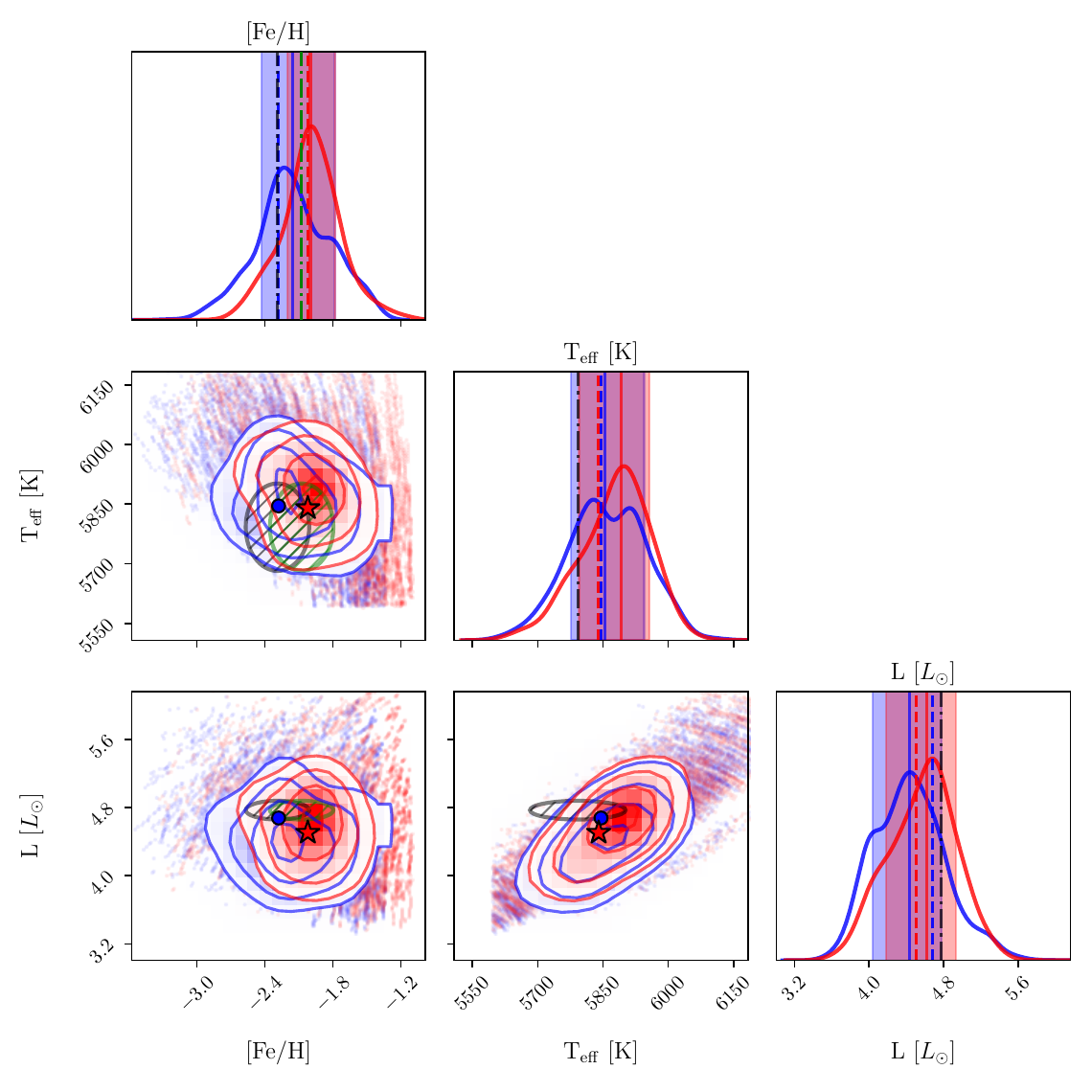}
    \caption{Same as \autoref{fig:corner_plot_global_params_HD_128279} except for HD 140283.  } 
    \label{fig:corner_plot_global_params_HD_140283}
\end{figure*}

\begin{figure*}
    \centering
    \includegraphics[width=0.49\textwidth]{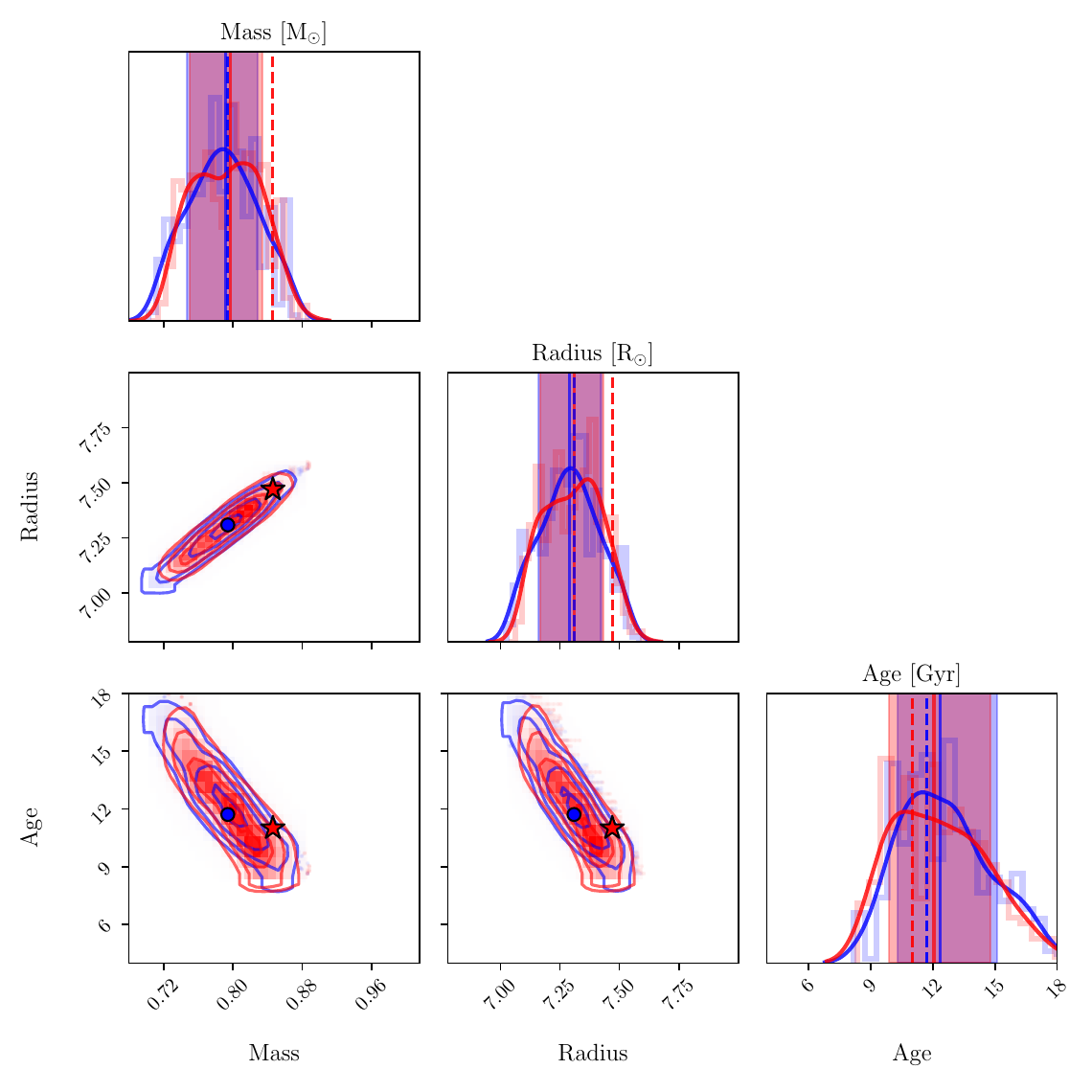}
    \includegraphics[width=0.49\textwidth]{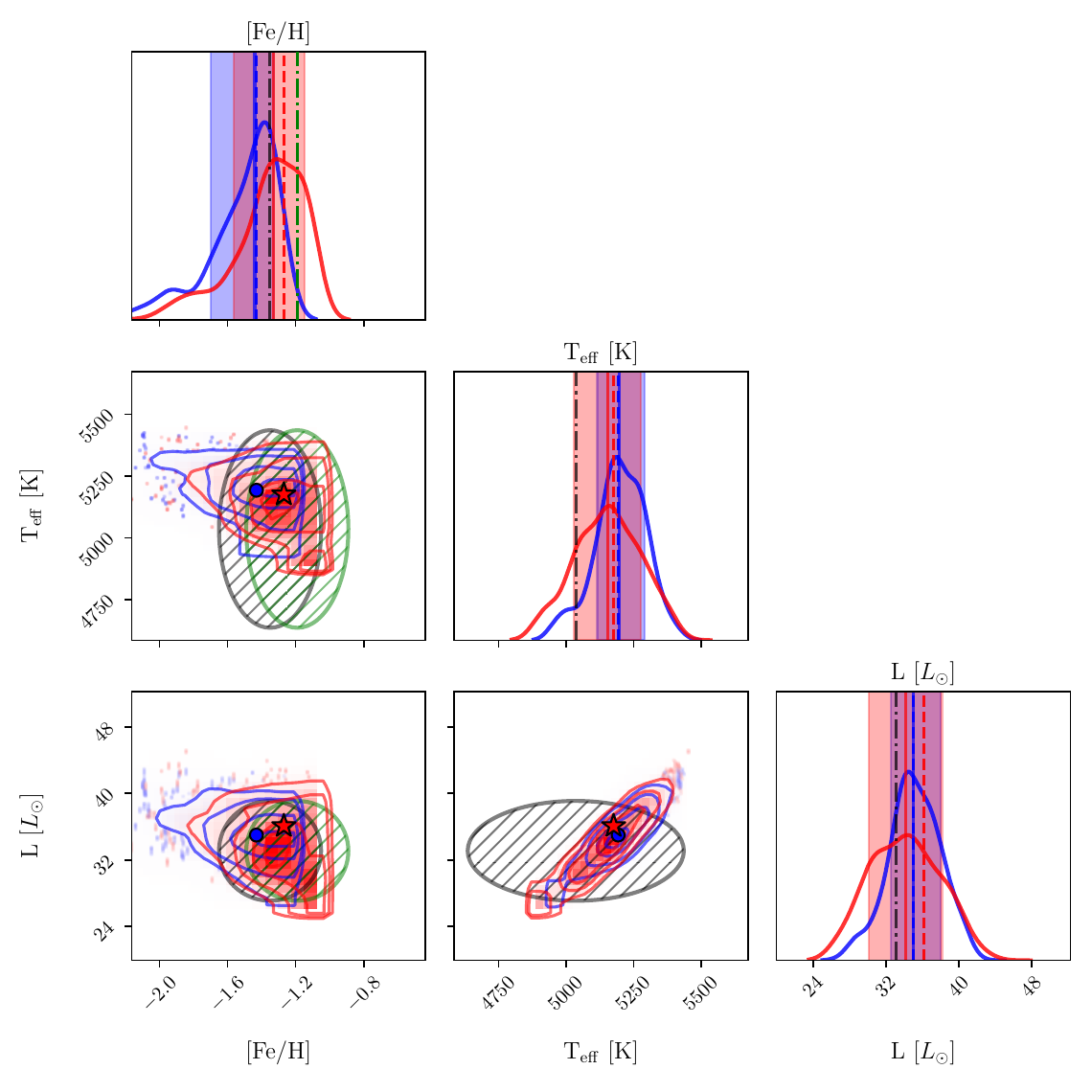}
    \caption{Same as \autoref{fig:corner_plot_global_params_HD_128279} except for HD 175305. } 
    \label{fig:corner_plot_global_params_HD_175305}
\end{figure*}

\begin{figure*}
    \centering
    \includegraphics[width=0.49\textwidth]{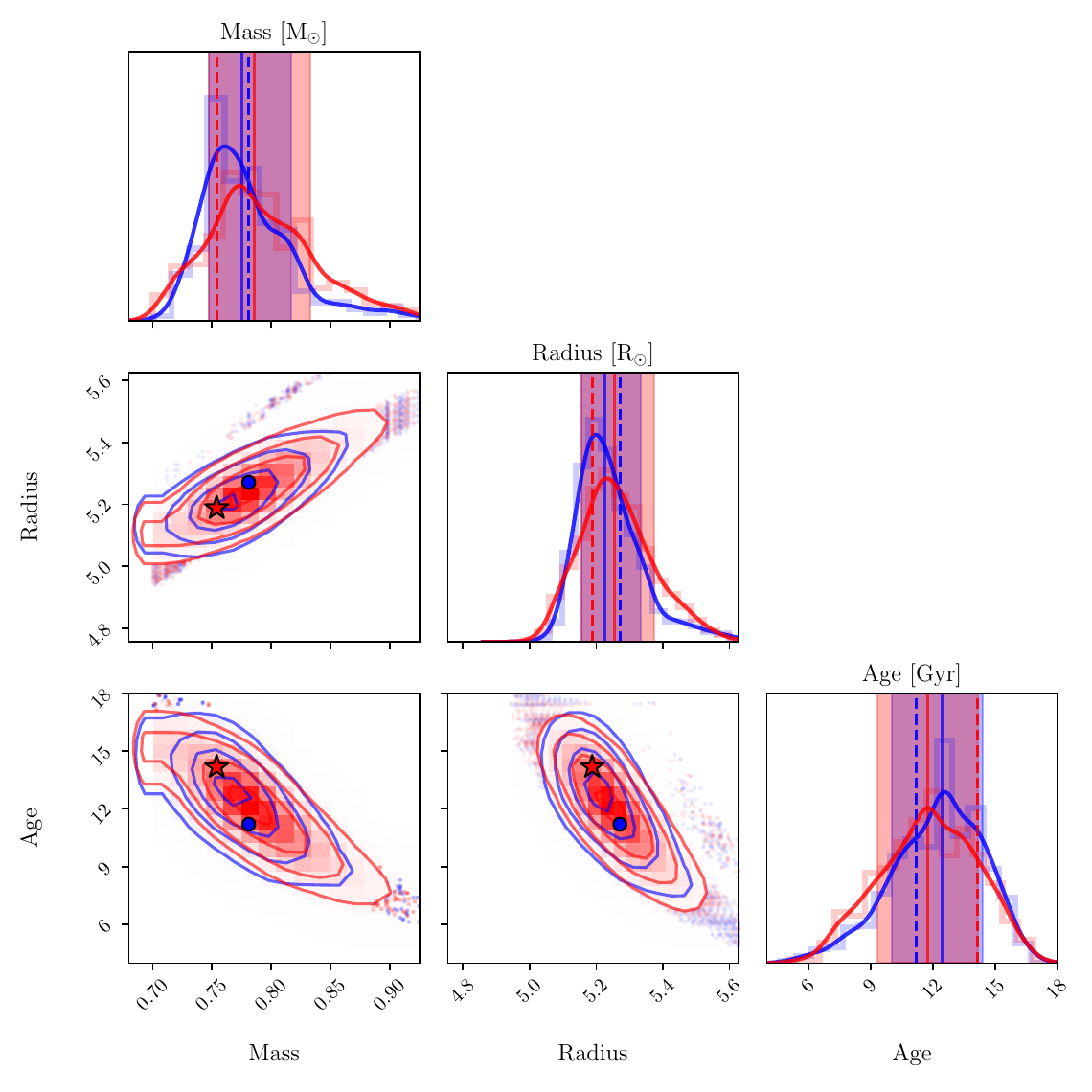}
    \includegraphics[width=0.49\textwidth]{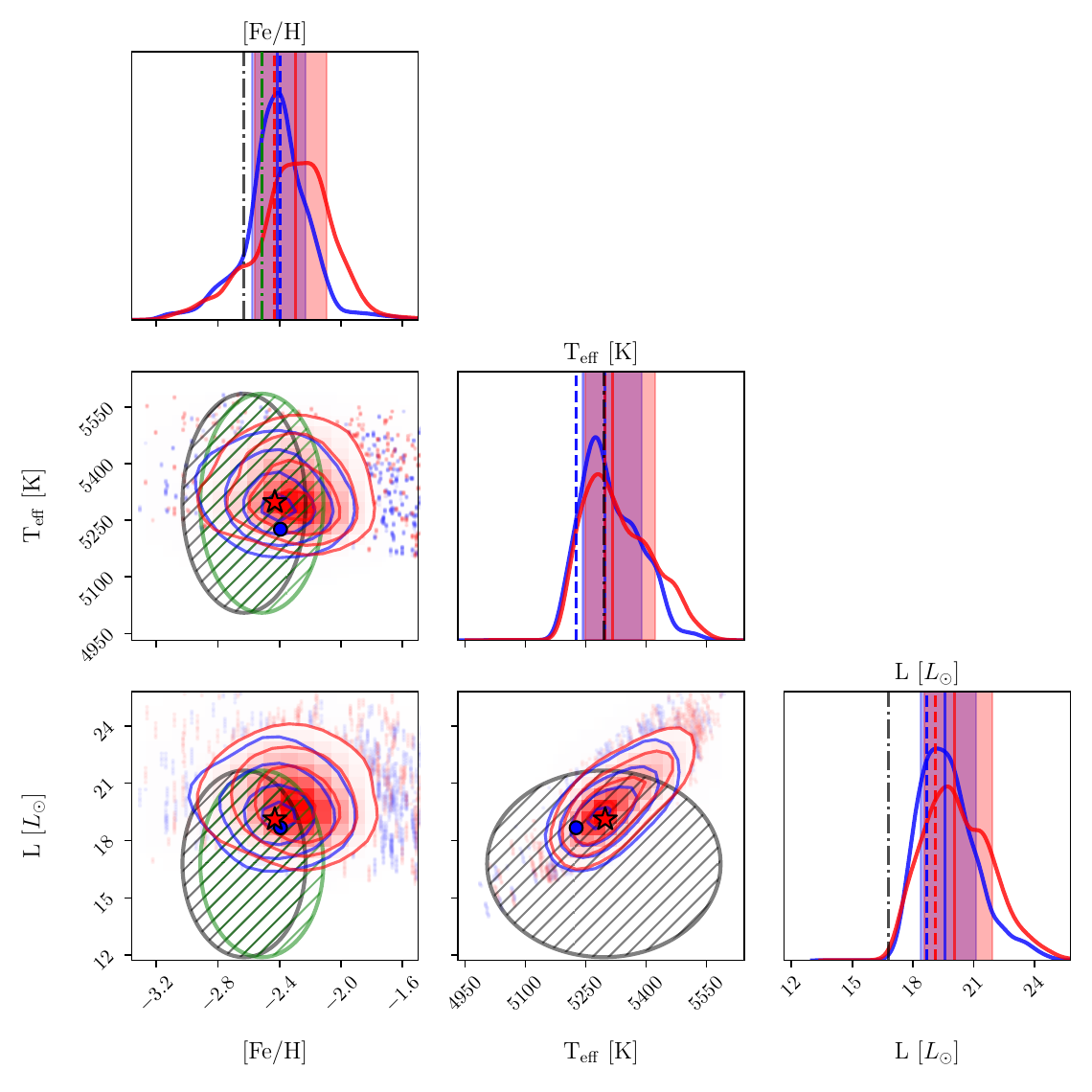}
    \caption{Same as \autoref{fig:corner_plot_global_params_HD_128279} except for KIC 4671239. } 
    \label{fig:corner_plot_global_params_KIC_4671239}
\end{figure*}

\begin{figure*}
    \centering
    \includegraphics[width=0.49\textwidth]{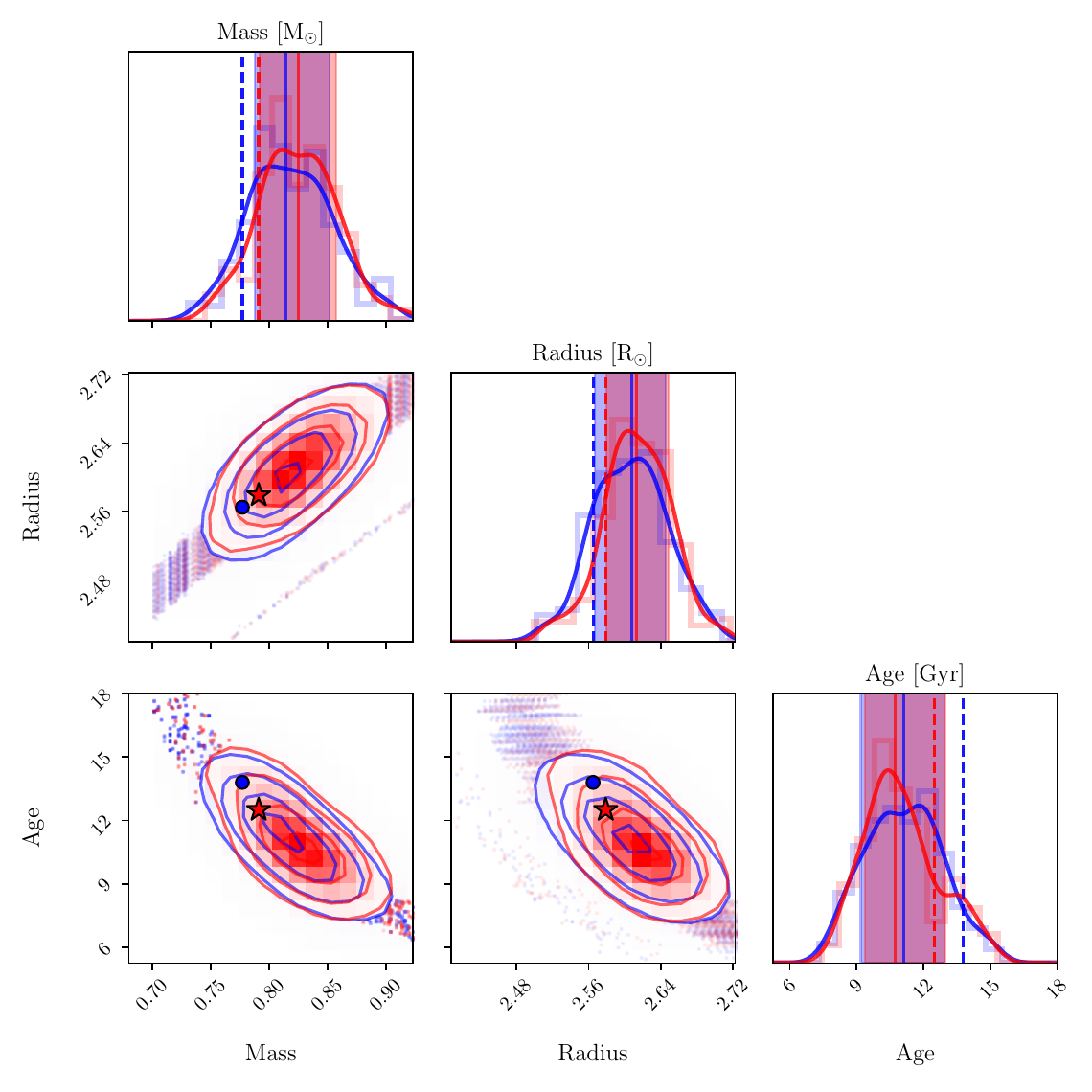}
    \includegraphics[width=0.49\textwidth]{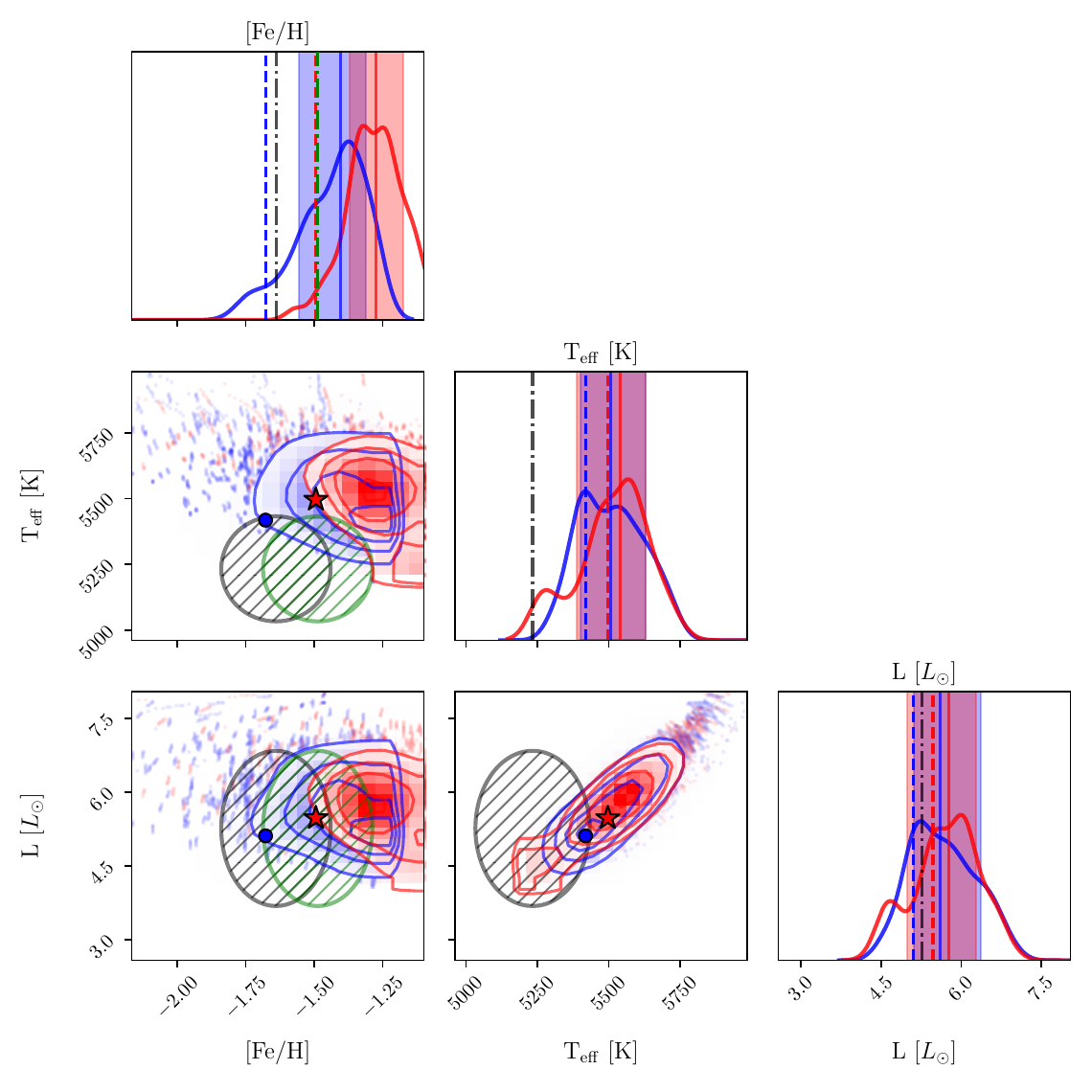}
    \caption{Same as \autoref{fig:corner_plot_global_params_HD_128279} except for KIC 7341231. } 
    \label{fig:corner_plot_global_params_KIC_7341231}
\end{figure*}

\begin{figure*}
    \centering
    \includegraphics[width=0.49\textwidth]{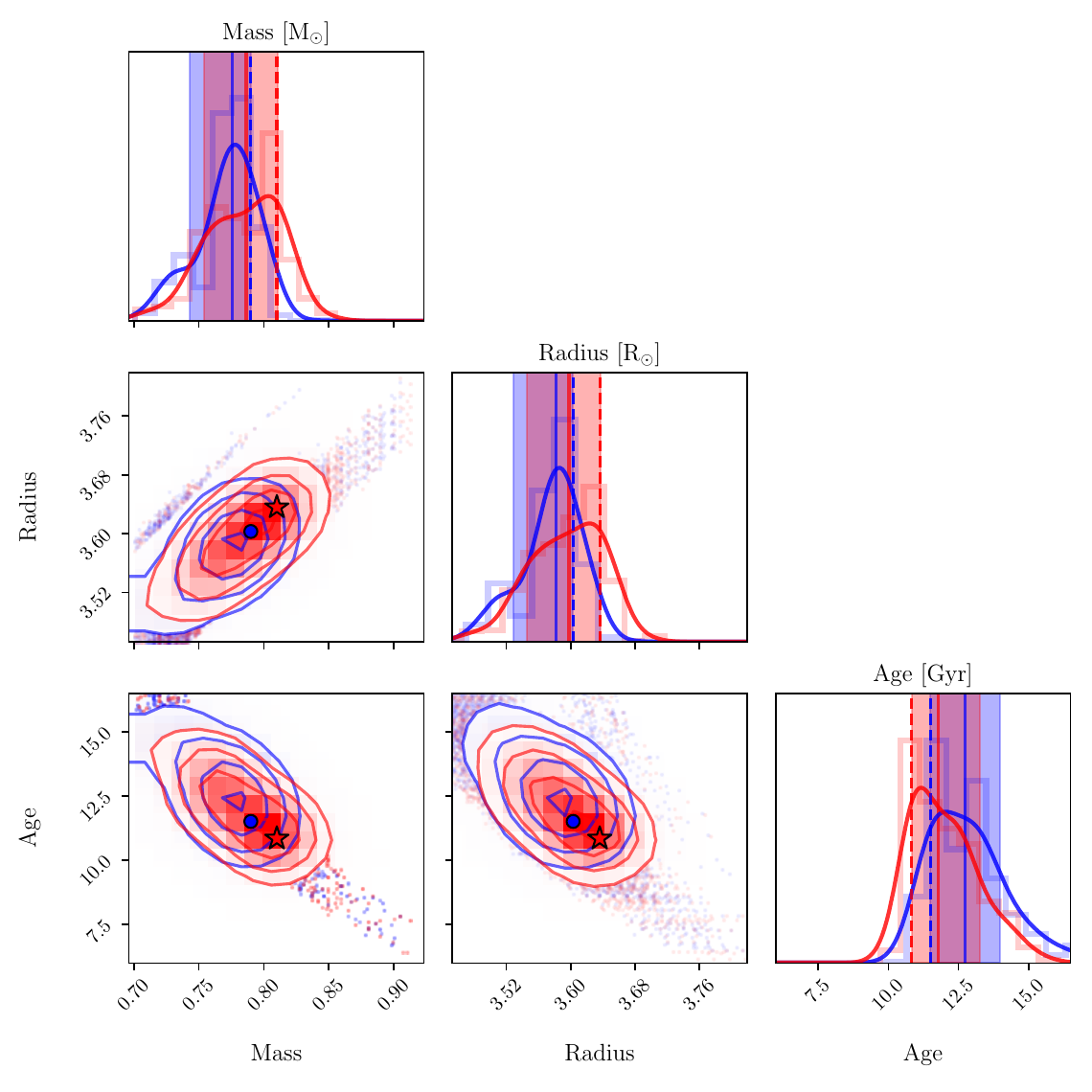}
    \includegraphics[width=0.49\textwidth]{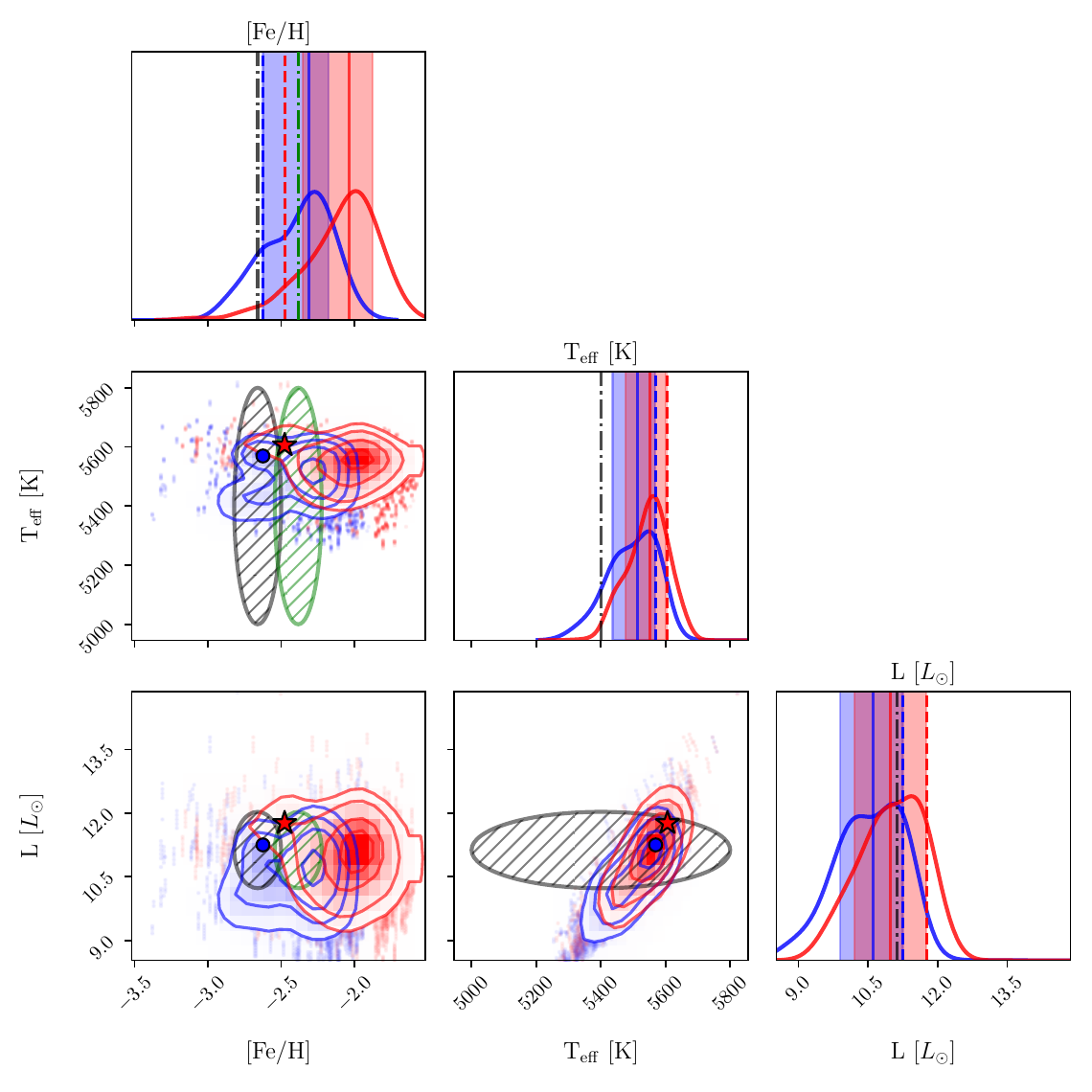}
    \caption{Same as \autoref{fig:corner_plot_global_params_HD_128279} except for KIC 8144907. } 
    \label{fig:corner_plot_global_params_KIC_8144907}
\end{figure*}

\begin{figure*}
    \centering
    \includegraphics[width=0.49\textwidth]{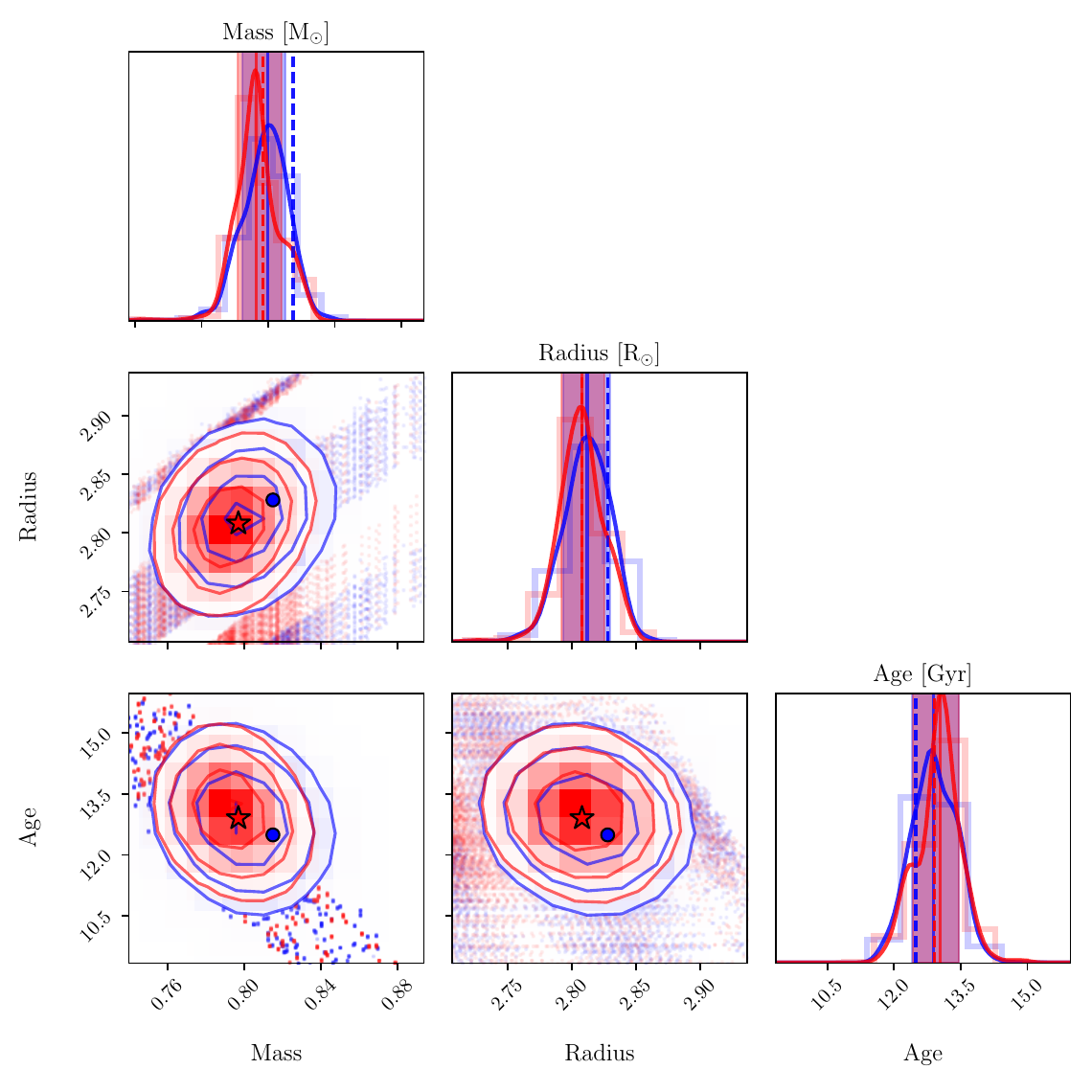}
    \includegraphics[width=0.49\textwidth]{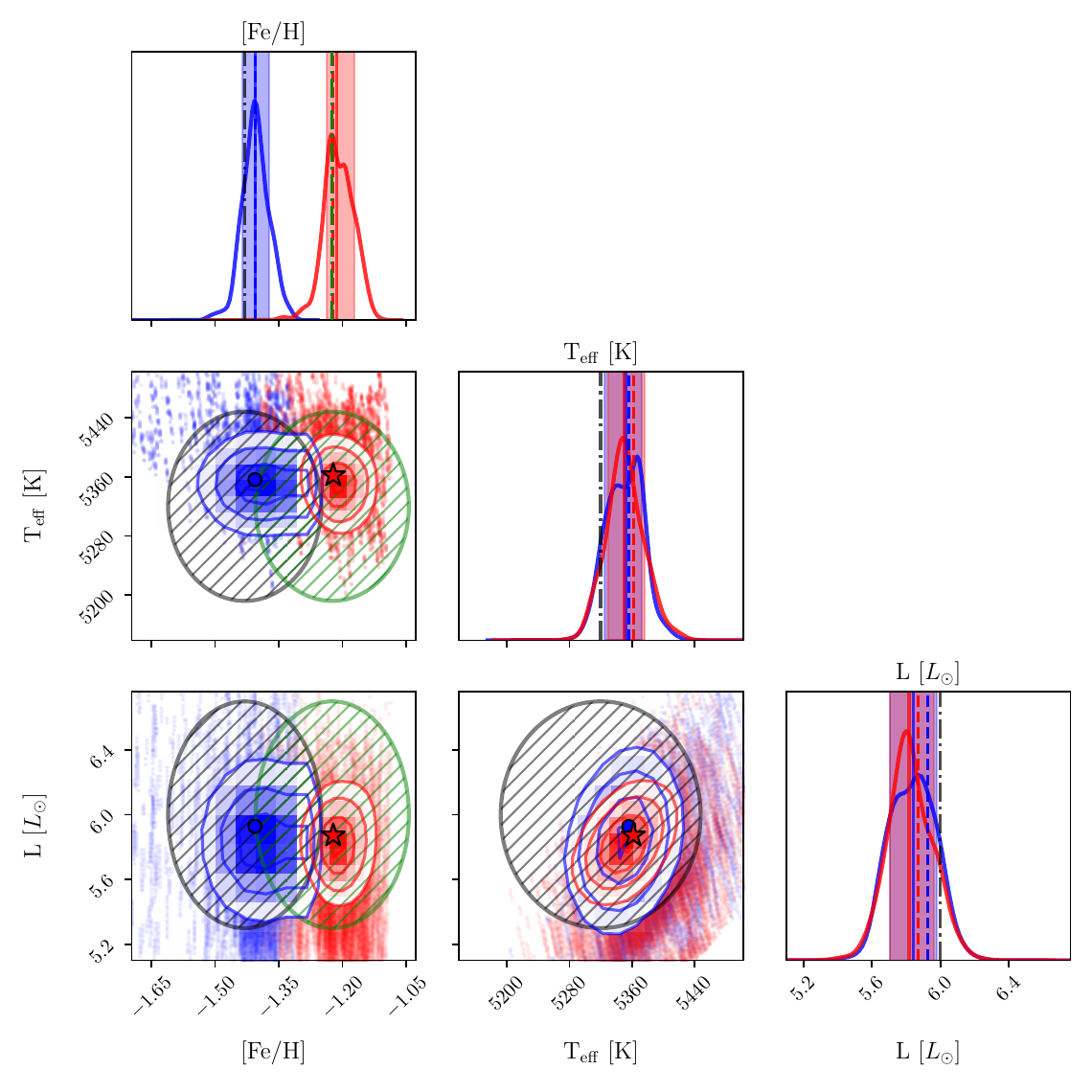}
    \caption{Same as \autoref{fig:corner_plot_global_params_HD_128279} except for $\nu$ Indi. } 
    \label{fig:corner_plot_global_params_nu_indi}
\end{figure*}

\autoref{fig:corner_plot_global_params_HD_128279}, 
\autoref{fig:corner_plot_global_params_HD_140283}, \autoref{fig:corner_plot_global_params_HD_175305}, \autoref{fig:corner_plot_global_params_KIC_4671239}, \autoref{fig:corner_plot_global_params_KIC_7341231}, \autoref{fig:corner_plot_global_params_KIC_8144907}, and\autoref{fig:corner_plot_global_params_nu_indi}, show the global parameter and spectroscopic parameter corner plots similar to \autoref{fig:corner_plot_global_params} and \autoref{fig:corner_plot_spec} for the other stars in our sample. See \autoref{sec:optimization} for a description of each corner plot's construction. 


\end{document}